\documentclass[aps,prd,nofootinbib,showkeys,preprint,floatfix]{article}

\pdfoutput=1

\usepackage{graphicx}
\usepackage{graphics}
\usepackage[mathscr]{eucal}
\usepackage{amssymb}
\usepackage{amsmath}
\usepackage{xspace}
\usepackage{listings}
\usepackage{ulem}
\usepackage{color}
\usepackage{jheppub}

\definecolor{darkgreen}{rgb}{0,0.5,0}

\newcommand{\vev}[0]{VEV\xspace}
\newcommand{\vevs}[0]{VEVs\xspace}
\newcommand{\DRbar}[0]{\ensuremath{\overline{\mathrm{DR}}}}
\newcommand{\gev}[0]{\text{ GeV}\xspace}

\newcommand{\Eq}[1]{eq.~(#1)\xspace}

\newcommand{\Tab}[0]{table\xspace}
\newcommand{\Fig}[0]{figure\xspace}

\newcommand{\eg}[0]{{e.g}.\xspace}

\newcommand{\ie}[0]{{i.e}.\xspace}

\newcommand{\Ref}[0]{Ref.\xspace}
\newcommand{\Refs}[0]{Refs.\xspace}
\newcommand{\Sec}[0]{section\xspace}
\def\gsim{\raise0.3ex\hbox{$\;>$\kern-0.75em\raise-1.1ex\hbox{$\sim\;$}}}

\newcommand{\sarah}[0]{\texttt{SARAH}\xspace}

\newcommand{\vevacious}[0]{\texttt{Vevacious}\xspace}
\newcommand{\vcs}[0]{\vevacious}

\newcommand{\cosmotransitions}[0]{\texttt{CosmoTransitions}\xspace}
\newcommand{\ct}[0]{\cosmotransitions}
\newcommand{\spheno}[0]{\texttt{SPheno}\xspace}
\newcommand{\softsusy}[0]{\texttt{SoftSUSY}\xspace}

\newcommand{\sL}[0]{\ensuremath{{\tilde{\tau}}_{L}}\xspace}
\newcommand{\sE}[0]{\ensuremath{{\tilde{\tau}}_{R}}\xspace}
\newcommand{\tsq}[0]{\ensuremath{\tilde{t}}\xspace}
\newcommand{\sQ}[0]{\ensuremath{{\tsq}_{L}}\xspace}
\newcommand{\sU}[0]{\ensuremath{{\tsq}_{R}}\xspace}
\newcommand{\vd}[1]{\ensuremath{v_{d}^{#1}}\xspace}
\newcommand{\vu}[1]{\ensuremath{v_{u}^{#1}}\xspace}
\newcommand{\vl}[1]{\ensuremath{v_{L3}^{#1}}\xspace}
\newcommand{\ve}[1]{\ensuremath{v_{E3}^{#1}}\xspace}
\newcommand{\vq}[1]{\ensuremath{v_{Q3}^{#1}}\xspace}
\newcommand{\vt}[1]{\ensuremath{v_{U3}^{#1}}\xspace}
\def\lsim{\raise0.3ex\hbox{$\;<$\kern-0.75em\raise-1.1ex\hbox{$\sim\;$}}}

\newcommand\ELIELout{\bgroup\markoverwith
{\textcolor{blue}{\rule[.5ex]{2pt}{0.4pt}}}\ULon}

\newcommand{\AddrBonn}{%
Bethe Center for Theoretical Physics \& Physikalisches Institut der
 Universit\"at Bonn, \\
53115 Bonn, Germany }

\newcommand{\AddrWur}{%
Institut f\"ur Theoretische Physik und Astronomie,
Universit\"at W\"urzburg\\
Am Hubland,
97074 W\"urzburg, Germany}

\title{Stability of the CMSSM against sfermion VEVs}

\author[a]{J.\ E.\ Camargo-Molina}
\author[a]{B.\ O'Leary}
\author[a]{W.\ Porod}
\author[b]{F.\ Staub}

\affiliation[a]{\AddrWur}
\affiliation[b]{\AddrBonn}

\emailAdd{jose.camargo@physik.uni-wuerzburg.de}
\emailAdd{ben.oleary@physik.uni-wuerzburg.de}
\emailAdd{porod@physik.uni-wuerzburg.de}
\emailAdd{fnstaub@th.physik.uni-bonn.de}


\preprint{Bonn-TH-2013-16}
\abstract{
The recent discovery of a Higgs boson by the LHC experiments has profound
 implications for supersymmetric models. In particular, in the context of
 restricted models, such as the supergravity-inspired constrained minimal
 supersymmetric standard model, one finds that preferred regions in parameter
 space have large soft supersymmetry-breaking trilinear couplings. This
 potentially gives rise to charge- and/or color-breaking minima besides those
 with the correct breaking of $SU(2)_L \times U(1)_Y$. We investigate the
 stability of parameter points in this model against tunneling to possible
 deeper color- and/or charge-breaking minima of the one-loop effective
 potential. We find that allowed regions of the parameter space with light staus
 or with light stops are seriously constrained by the requirement that there are
 no deeper minima, and the parameter space is still quite constrained even by
 the less strict requirement that the tunneling time out of the normal
 electroweak-symmetry-breaking vacuum is more than a fifth of the age of the
 known Universe. We also find that ``thumb rule'' conditions on Lagrangian
 parameters based on specific directions in the tree-level potential are of
 limited use.
}

\begin{document}
\maketitle
\flushbottom

\section{Introduction}
\label{sec:intro}

The standard model of particle physics (SM)
 \cite{Glashow:1961tr, Weinberg:1967tq, Salam:1968rm} has proven itself to be an
 extremely good description of particle physics all the way up to the
 tera-electron-Volt scale. The interpretation of the bosonic resonance at
 $125 \gev$ recently discovered at the Large Hadron Collider (LHC)
 \cite{Aad:2012tfa,Chatrchyan:2012ufa} as the Higgs boson of the SM completes
 the picture, and allows us to probe the mechanism of the spontaneous breaking
 of gauge symmetries.

A particular issue, however, of the Higgs boson is that its mass, determining
 the electroweak scale, is many orders of magnitude smaller than the ``natural''
 value of the order of the Planck scale. This, along with the fact that quantum
 corrections to the mass of the Higgs boson are typically of the scale of the
 heaviest particles which interact with it, leads to model builders attempting
 to resolve this ``hierarchy problem''.

One popular way of ameliorating the hierarchy problem is to promote the SM to a
 supersymmetric theory, such as the minimal supersymmetric standard model (MSSM)
 (see \cite{Martin:1997ns} for a review). The MSSM contains all the particles
 and interactions of the SM as a subset, and the interactions of the
 supersymmetric partners are related to the SM interactions.

Usually one postulates a conserved parity known as $R$-parity
 \cite{Farrar:1978xj} to avoid baryon- and lepton-number violating interactions
 which would be otherwise allowed by gauge symmetries and supersymmetry. The
 conservation of this parity, which we take here as part of the definition of
 the MSSM, implies the stability of the lightest supersymmetric partner (LSP),
 which, if uncharged under $SU(3)_{c} \times U(1)_{EM}$, is a candidate particle
 to explain the observed dark matter of the Universe.
 
In principle, the MSSM has {\it less} parameters than the SM, since the quartic
 coupling of the Higgs boson is actually a function of the gauge couplings.
 However, the mechanism of supersymmetry (SUSY) breaking must introduce more
 parameters, and agnosticism of the exact mechanism leads to the common practice
 of parametrizing the mechanism by adding {\it soft SUSY-breaking terms} to the
 Lagrangian density. The number of parameters specifying the full set of soft
 SUSY-breaking terms allowed in the MSSM is rather large, namely 105
 \cite{Martin:1997ns}, so often they are taken to be related at a specific
 scale. One of the simplest and most popular proposals is the
 minimal-supergravity-inspired constrained MSSM (CMSSM), in which all the soft
 SUSY-breaking scalar mass-squared terms are taken to be equal to $M_{0}^{2}$
 at the scale $M_{GUT}$ where the gauge couplings unify, assuming that somehow
 the MSSM is a part of a grand unified theory (GUT). In addition, the soft
 SUSY-breaking mass terms for the fermionic partners of the gauge bosons are
 also taken to unify at $M_{GUT}$ with a common value $M_{1/2}$. Finally a third
 GUT-scale common value is defined: $A_{0}$, a common trilinear scalar
 interaction coupling (multiplied by the corresponding Yukawa couplings). In
 principle, the $\mu$ parameter coupling the Higgs doublets and the
 corresponding soft SUSY-breaking term $B_{\mu}$ could also be defined at the
 GUT scale, but for practical reasons they are taken to be engineered by the
 requirement of correct electroweak symmetry breaking at a given scale, often
 the geometric mean of the masses of the two stops, the scalar partners of the
 top quark, referred to as the SUSY scale. The values of $|\mu|$ and $B_{\mu}$
 are fixed by requiring that the mass of the $Z$ boson is correct along with
 defining the ratio $\tan \beta$ of the {\it vacuum expectation values} (\vevs)
 \vd{} and \vu{} of the neutral components of the two Higgs doublets, and the
 sign of $\mu$ is given as a final input\footnote{However, it has been noted in
 \Ref~\cite{Allanach:2013cda} that defining a Lagrangian partially at different
 scales is ambiguous and in the case of the CMSSM, different values of $\mu$
 and $B_{\mu}$ at the GUT scale can lead to the same $m_{Z}$ and $\tan \beta$
 at the SUSY scale while keeping $M_{0}, M_{1/2}$, and $A_{0}$ the same. This
 is indeed interesting, but beyond the scope of this work, where we identify
 CCB minima of SUSY-scale Lagrangians  which were generated by CMSSM
 conditions.}.

The constraints of the CMSSM broadly lead to fixed ratios of the gaugino masses
 at the SUSY scale, and groupings of the squark masses together and the slepton
 masses together. The size of the gaugino masses and the masses of the groupings
 of sfermions relative to the gauginos are controlled by $M_{1/2}$ and $M_{0}$
 respectively, and the only remaining freedom to change this spectrum lies in
 tuning $A_{0}$ and $\tan\beta$ to separate out the third-generation sfermions
 from the other generations.

The interplay between this handful of parameters is usually enough to explain
 many observations. For instance, requiring that the relic density of dark
 matter is within the uncertainties of the observed value might seem to
 significantly constrain the allowed parameter space \cite{Olive:1989jg}.
 However, if one fixes one or two of the CMSSM parameters, there is typically a
 range of the other parameters where the relic density is compatible with
 observations \cite{Ellis:1998kh}. One particular region is known as the
 ``stau co-annihilation region'' \cite{Ellis:1998kh}, where the stau is
 marginally heavier than the lightest neutralino, which is the LSP, and freezes
 out only slightly earlier, allowing more annihilation before the neutralino
 freeze-out.

At the same time a mass of $125 \gev$ for the lightest MSSM Higgs boson is
 difficult to achieve without at least one stop with a multi-TeV mass
 \cite{Draper:2011aa, Heinemeyer:2011aa, Brummer:2012ns, Djouadi:2013vqa,%
 Arbey:2012bp, Arganda:2012qp}. One way to achieve this is to have a
 sufficiently high $M_{0}$ so that the stop mass is large enough, and a
 sufficiently high $M_{1/2}$ to bring the mass of the lightest neutralino up to
 just below the lightest stau mass (though for sufficiently high masses, the
 stau co-annihilation mechanism cannot reduce the relic density to the observed
 value, even for a vanishing mass difference between the stau and the lightest
 neutralino \cite{Ellis:1998kh, Citron:2012fg}). However, this region of
 parameter space has rather bleak prospects for the discovery of sparticles. The
 only other way that this can be achieved within the CMSSM is through a large
 $A_{0}$, inducing a large splitting between the two mass eigenstates of both
 the staus and the stops. This allows the loop corrections to the Higgs mass to
 be large through both the existence of a heavy stop and the large splitting
 between the stop mass eigenstates, while allowing at least some potentially
 LHC-accessible sparticles \cite{Bechtle:2012zk}.

The presence of many additional scalar partners for the SM fermions raises the
 question of whether they too could develop \vevs. If, for example, the
 potential were such that stops would develop non-zero \vevs, then that would
 be disastrous as these \vevs would spontaneously break $SU(3)_{c}$ and
 $U(1)_{EM}$! Unfortunately until recently it was quite impractical to search
 for other vacua to see whether the desired vacuum is stable, or whether there
 are charge- and/or color-breaking (CCB) minima.

Since there are many different possibilities for vacua in the MSSM,
 spontaneously breaking any or all of the gauge symmetries, we denote the
 vacuum that should describe the Universe in which we live as the
 desired-symmetry-breaking (DSB) vacuum.

Some of the earliest work investigated the directions of the tree-level
 potential where the quartic terms vanish, as soft SUSY-breaking terms could
 lead to the potential being unbounded from below in these directions or to CCB
 minima deeper than the DSB vacuum developing along these directions
 \cite{Nilles:1982dy, AlvarezGaume:1983gj, Derendinger:1983bz, Claudson:1983et,%
 Kounnas:1983td, Drees:1985ie, Gunion:1987qv, Komatsu:1988mt, Langacker:1994bc,%
 Casas:1995pd, Casas:1996de}. Moreover, studies have been performed on the 
 tunneling time between different vacua \cite{Kusenko:1996jn}. Certain
 conditions relating the trilinear parameters with the mass-squared parameters
 have been obtained which ensured that no deeper minimum could develop
 {\it along a line where the scalar fields have values in fixed ratios to each
 other}. However, it has been known for many years that it is only very special
 parameter points where this condition would be sufficient to forbid undesired
 minima, and that in general even at tree level it is very difficult to ensure
 that there are no undesired minima deeper than the desired minimum
 \cite{Abel:1998ie}. Moreover, despite various claims in the literature
 \cite{PhysRevD.48.4352, Casas:1995pd}, loop corrections are important: in
 \cite{Bordner:1995fh}, a numerical minimization of the one-loop effective
 potential including the top-quark Yukawa contributions has been performed
 demonstrating the importance of the corrections, and it was noted in
 \cite{Ferreira:2000hg} that loop corrections could change the ordering of which
 minimum is deepest.

It is possible that CCB minima could be tolerated
 \cite{Riotto:1995am, Kusenko:1996xt, Kusenko:1996jn}, if the Universe would
 have fallen naturally into the false DSB vacuum as the cosmological temperature
 decreased, and if the lifetime of this vacuum for tunnelling into the true CCB
 vacuum is much longer than the present age of the Universe. Whether the DSB
 vacuum is in fact preferred by cosmology depends, in particular, on the scalar
 masses-squared generated during inflation. If these masses-squared are
 positive and of the order of the square of the Hubble parameter, the `more
 symmetric' DSB vacuum is favored. On the other hand, if these are negative, the
 Universe would remain trapped in the true CCB vacuum \cite{Falk:1996zt}. A
 detailed discussion on these issues including the case of negative $M^2_0$ at
 $M_{GUT}$ and higher dimensional operators can be found in \cite{Ellis:2008mc}. 

In the last few years there has been much progress in the field of determining
 the global minimum of the potential of a quantum field theory. In particular,
 the global minimum of renormalizable tree-level potentials (and other
 potentials of a purely polynomial form) can now be found deterministically
 with methods such as the Gr{\"{o}}ber basis method
 \cite{Maniatis:2006jd, Gray:2008zs} or the homotopy continutation method
 \cite{Huang199577, sommesenumerical, li2003numerical}.

Recently the power of the homotopy continuation method for finding tree-level
 minima combined with gradient-based minimization with loop corrections has
 been combined in the publicly-available code \vcs\
 \cite{Camargo-Molina:2013qva}. We use this tool to investigate regions of the
 CMSSM which, despite having local minima with the desired breaking of
 $SU(2)_{L} \times U(1)_{Y}$ to $U(1)_{EM}$ while preserving $SU(3)_{c}$, might
 have global minima with a different breaking of the gauge symmetries. This
 allows us to update existing studies
 \cite{Baer:1996jn, Strumia:1996pr, Ferreira:2000hg, Cerdeno:2003yt} by
 calculating {\it all} the tree-level extrema and the complete one-loop
 effective potential in the neighborhood of these extrema. This also allows us
 to check to what extent the existing rules are useful at all. This is
 particularly important in view of the fact that the explanation of the observed
 Higgs mass requires special parameter combinations.

This paper is organized as follows: in \Sec~\ref{sec:rules} we collect the
 existing analytical approximations to determine color- and/or
 charge-breaking minima. In \Sec~\ref{sec:method}, we briefly explain the tools
 that we used to generate the spectra of CMSSM points and to evaluate the
 stability of such points against undesired vacua. Then we consider how robust
 our results are against which scalars are allowed non-zero \vevs and against
 variations in scale, how dependent they are on loop corrections, and how the
 results might depend on the precise values of the Lagrangian parameters as
 evaluated by different spectrum generators; we also examine the usefulness of
 the tree-level conditions mentioned above. In \Sec~\ref{sec:constraining}, we
 investigate part of the stau co-annihilation region to demonstrate that its
 parameter points generally are only metastable, and we also demonstrate that it
 is difficult to get light stops in the CMSSM without rapid tunneling to CCB
 vacua. In addition, we show the stability of regions compatible with the
 measured mass of the Higgs boson, and further demonstrate the irrelevance of
 the ``thumb rules'' to the parameter space regions of interest.

\section{Analytical approximations for limits of charge- and color-breaking
 minima}
\label{sec:rules}

The tree-level scalar potential, $V^{\text{tree}}$ of the MSSM consists of
 three parts
\begin{equation}
 V^{\text{tree}} = V_D + V_F + V_\text{soft}
\end{equation}
 with
\begin{eqnarray}
 V_D &=&
 \frac{1}{2}\sum_a g_a^2 \left(\sum_i \phi^\dagger_i T^a \phi_i\right)^2\\
 V_F &=& \sum_i \left| \frac{\partial W}{\partial \phi_i}\right|^2 \\
 V_\text{soft}
 &=& \sum_i m^2_i  \left| \phi_i\right|^2
 + \left( B_\mu H_d H_u +\text{H.c.}\right) \nonumber \\
 &+& \sum_{\alpha = \text{generations}} \left(
 A_{u_\alpha} Y_{u_\alpha} H_u \tilde Q_\alpha \tilde u^*_{R,\alpha} + 
 A_{d_\alpha} Y_{d_\alpha}  \tilde Q_\alpha H_d \tilde d^*_{R,\alpha} +
 A_{l_\alpha} Y_{l_\alpha}  \tilde L_\alpha H_d \tilde e^*_{R,\alpha}
 + \text{H.c.}\right)
\end{eqnarray}
 where we have neglected flavor violation for simplicity. At the minimum the
 scalar fields take constant values. As a guide to aid in following the
 discussion on formulae based on $V^{\text{tree}}$, we show the case
 $H_{d} = \vd{}/\sqrt{2}, H_{u} = \vu{}/\sqrt{2}, \sL = \vl{}/\sqrt{2},
 \sE = \ve{}/\sqrt{2}$, all other scalars set to 0:
\begin{eqnarray}
V^{\text{tree}}_{H_{d}, H_{u}, \sL, \sE}
 & = & \frac{1}{32}
 \left( g_{1}^{2} ( \vd{2} - \vu{2} + \vl{2} - 2 \ve{2} )^{2}
                    + g_{2}^{2} ( \vd{2} - \vu{2} - \vl{2} )^{2} \right)
 \nonumber \\
 & & + \frac{1}{4} Y_{\tau}^{2} \left( \vd{2} \vl{2}
     + \vd{2} \ve{2} + \vl{2} \ve{2} \right)
     + \frac{ Y_{\tau} }{ \sqrt{2} } \vl{} \ve{} \left( A_{\tau} \vd{}
     - \mu \vu{} \right)
 \nonumber \\ 
 & & - B_{\mu} \vd{} \vu{}
     + \frac{1}{2} \left( | \mu |^{2} ( \vd{2} + \vu{2} )
     + m_{H_{d}}^{2} \vd{2} + m_{H_{u}}^{2} \vu{2}
     + m_{\sL}^{2} \vl{2} + m_{\sE}^{2} \ve{2} \right)
\label{eq:tree_Higgs_and_stau_potential}
\end{eqnarray}

The main focus of previous studies was to derive analytical conditions
 involving relevant parameters that would allow one to identify regions in
 parameter space without color- or charge-breaking minima. This led to a
 collection of ``thumb rules'' which do well only for specific extreme cases.

The most widespread of these is derived by making assumptions about the ratios
 of \vevs at CCB minima. Keeping the scalar fields at values in constant ratios
 to each other defines rays in field configuration space. An example that we
 shall consider in a moment would be to take the stau fields to be equal to the
 down-type Higgs field: $H_{d} = \sL = \sE$, and all other fields held to zero.
 Along such rays, the potential can be written as a function of a single
 variable, such as the normalized magnitude of the \vevs. The key observation
 underpinning the widespread thumb rules is that the $D$-term contributions to
 the potential vanish along certain rays, removing their ``stabilizing''
 influence (as they are always positive for non-zero values of the scalar
 fields). Hence one might suspect that undesired minima would lie in such
 regions; so that along these rays, one can solve the one-dimensional
 minimization condition (the $F$-terms ensure that the potential is still
 bounded from below), and compare the minimum along this ray to the DSB
 minimum.

Along a ray where the $D$-terms vanish, one can cast the tree-level potential
 in the form
\begin{equation}
 V^{\text{tree}}
 = \frac{ m^{2} }{2} v^{2} - \frac{ Y A }{3} v^{3} + \frac{Y^{2}}{4} v^{4}
\label{eq:useless_ray_form}
\end{equation}
 where $v$ is the (scaled) length of the vector of field values.
 This is minimized when
\begin{equation}
 v = \frac{ A + \sqrt{ A^{2} - 4 m^2 } }{ 2 Y }
\end{equation}
 (or when $v = 0$, where the tree-level potential is zero). Since the desired
 normal EWSB vacuum has a negative value for the tree-level potential, it is
 then sufficient to ensure that the potential along this ray is never negative.
 Plugging the above value of $v$ into $V^{\text{tree}}$ and demanding that the
 potential is positive independent of the sign of $A$ shows that the dependence
 on $Y$ is just an overall factor, and that the potential is positive at this
 field configuration regardless of the value of $Y$ as long as the condition
\begin{equation}
 m^{2} > \frac{2}{9} A^{2}
\label{eq:useless_generic_condition}
\end{equation}
 is satisfied, which then unpacks into various conditions, depending on which
 rays are taken. 

One of the simplest rays along which the $D$-terms of
 \Eq{\ref{eq:tree_Higgs_and_stau_potential}} vanish is the direction where
 $H_{u}$ is taken to be $0$, and $H_{d} = \sL = \sE = 3^{-1/4} v$, where the
 factor of $3^{-1/4}$ keeps the quartic term correctly normalized when casting
 the potential in the form of \Eq{\ref{eq:useless_ray_form}} and thus
 $m^{2} = ( m^{2}_{H_{d}} + |\mu|^{2} + m^{2}_{\sL} + m^{2}_{\sE} ) / \sqrt{3}$
 and $A = 2^{-1/2} 3^{1/4} A_{\tau}$. This leads to a widely-used condition
 \cite{Nilles:1982dy, AlvarezGaume:1983gj, Derendinger:1983bz, Claudson:1983et,%
 Kounnas:1983td}
\begin{equation}
A_{\tau}^{2} < 3 ( m_{H_{d}}^{2} + |\mu|^{2} + m_{\sL}^{2} + m_{\sE}^{2} )
\label{eq:useless_stau_condition}
\end{equation}
 and the analogous condition for stop \vevs is
\begin{equation}
A_{t}^{2} < 3 ( m_{H_{u}}^{2} + |\mu|^{2} + m_{\sQ}^{2} + m_{\sU}^{2} ) \,.
\label{eq:useless_stop_condition}
\end{equation}
In \Ref~\cite{Casas:1995pd} an improved set of conditions were given, which for
 example for the first generation take a similar form as above, \eg
\begin{equation}
 A_{u}^{2} < 3 ( m_{H_{u}}^{2} + m_{\tilde u_L}^{2} + m_{\tilde u_R}^{2} ) \,.
\label{eq:useless_sup_condition}
\end{equation}
 As here no large Yukawa couplings are involved, this can readily be expressed 
 using the high scale parameters of the CMSSM \cite{Ellwanger:1999bv}
\begin{equation}
 ( A_{0} - 0.5 M_{1/2} )^{2} < 9 M_{0}^{2} + 2.67 M_{1/2}^{2}
\label{eq:useless_GUT_condition}
\end{equation}
 This condition was derived explicitly for the \vevs of the sfermions of the
 first two generations as it relies on the Yukawa couplings being much smaller
 than the gauge couplings. This is manifestly not the case for the stop sector,
 nor for the stau sector for parameter points with large $\tan\beta$, and was
 noted as such. However, it is worth examining how well such conditions do for
 the third generation nevertheless, given the lack of any other simple
 expressions for stop or stau \vevs. Henceforth when we refer to
 condition~(\ref{eq:useless_sup_condition}), we take the parameters for the
 {\it third} generation of up-squark, \ie the stop, as opposed to the first
 generation:
 $A_{t}^{2} < 3 ( m_{H_{u}}^{2} + m_{\tilde t_L}^{2} + m_{\tilde t_R}^{2} )$.

Since we will be plotting the lines between points which satisfy these
 conditions quite frequently, we note here the color scheme that we use. The
 solid colored lines correspond to the dividing lines between parameter points
 that satisfy or fail the conditions as following:
 condition~(\ref{eq:useless_stau_condition}) by purple,
 condition~(\ref{eq:useless_stop_condition}) by orange,
 condition~(\ref{eq:useless_sup_condition}) by dark red, and
 condition~(\ref{eq:useless_GUT_condition}) by brown. We do not plot any of
 these lines in white regions in our figures, where no spectra could be
 calculated for the DSB minimum anyway, due to the presence of tachyons.

For the third generation, the expressions are considerably more complicated, and
 a numerical approach is usually necessary to get reasonable results. An
 algorithm to extract CCB minima of the tree-level potential assuming vanishing
 (color) $D$-terms and either stau or stop \vevs (but not both at once) was
 presented in \Ref~\cite{Casas:1995pd}. The authors assumed that
 $\tan\beta=\frac{v_u}{v_d}>1$ holds even at the CCB minimum, while our ansatz
 is more general and we find that often $\tan\beta<1$ holds at the global
 minimum for points with stop \vevs. We will also later compare our
 numerical results with the analytical constraints summarized above.

Another algorithm was presented in \cite{LeMouel:2001ym} for the limiting case
 of $\tan\beta \to \infty$. The algorithm only goes so far as to give a very
 conservative upper bound on $|A_{t}|$ to ensure that there is no minimum of the
 tree-level potential when the only non-zero \vevs are those for $H_{u}$, $\sQ$,
 and $\sU$. Since, as will be seen, the limit of large $\tan\beta$ is
 characterized by {\it stau} \vevs rather than stop \vevs within the CMSSM, the
 condition is not very helpful, as will be shown. Rather than implement the full
 numerical algorithm, however, we plot lines dividing points which fail
\begin{equation}
 A_{t}^{2} < ( 0.65 - 0.85 )^{2} ( 3 ( m_{{\tilde{t}}_{1}}^{2}
 + m_{{\tilde{t}}_{2}}^{2} + 2 m_{t}^{2} ) )
\label{eq:stop_large_tb_condition}
\end{equation}
 from those which pass in dark blue, where we have chosen $0.65$ from the point
 where the CCB condition diverges from the tachyonic stop line in \Fig~1 of
 \cite{LeMouel:2001ym} to $0.85$ as being close to the maximal allowed value
 from the optimal bound therein.

It has already been reported in the literature that these conditions are
 neither necessary nor sufficient to ensure that the desired minimum is the
 global minimum, even at tree level \cite{Gunion:1987qv, Ferreira:2000hg}. Note
 that the condition (\ref{eq:useless_generic_condition}) is {\it sufficient} to
 guarantee that {\it along this ray}, there is not a deeper undesired minimum,
 but not {\it necessary}. However ensuring that there is no deeper minimum
 along this ray is {\it necessary} but {\it not sufficient} to guarantee that
 there is no deeper minimum than the DSB minimum. The interplay of this logic
 is though that this condition is {\it neither necessary nor sufficient} to
 ensure that there is not a deeper undesired minimum for a field configuration
 which does not lie on this ray. There could be an undesired minimum along this
 ray that has a negative tree-level potential value, yet is not as negative as
 that of the DSB vacuum, or there could be a deeper undesired minimum which
 does not lie on the ray even though there is no point on the ray itself that
 has a negative tree-level potential energy.

Recently, the hints for an enhanced di-photon decay rate have revived the
 interest in charge-breaking minima since the only possibility to explain this
 enhancement in the MSSM which does not affect the other decay channels would be
 very light staus \cite{Carena:2012gp,Carena:2012mw}. In this context, new
 checks for charge-breaking minima due to stau \vevs have been derived or
 revived \cite{Hisano:2010re,Kitahara:2013lfa}:
\begin{eqnarray}
 \mu \tan \beta
 & < & 213.5 \sqrt{ m_{\sL} m_{\sE} } - 17.0 ( m_{\sL} + m_{\sE} )
 + 4.52 \times 10^{-2} {\gev}^{-1} ( m_{\sL} - m_{\sE} )^{2}
\nonumber \\
 &   & - 1.30 \times 10^{4} \gev
\label{eq:older_numeric_stau_condition} \\
 | ( Y_{\tau} v_{u} \mu ) / ( \sqrt{2} m_{\tau} ) |
 & < & 56.9 \sqrt{ m_{\sL} m_{\sE} } + 57.1 ( m_{\sL} + 1.03 m_{\sE} )
 - 1.28 \times 10^{4} \gev
\nonumber \\
 &   & + \frac{ 1.67 \times 10^{6} \gev^{2} }{m_{\sL} + m_{\sE}}
 - 6.41 \times 10^{6} \gev^{3} ( \frac{1}{m_{\sL}^{2}} 
 + \frac{0.983}{m_{\sE}^{2}} )
\label{eq:newer_numeric_stau_condition} 
\end{eqnarray}
 where $m_{\tau}$ is the pole mass of the fermionic $\tau$ lepton. The
 condition given by (\ref{eq:older_numeric_stau_condition}) was obtained by a
 numerical fit of the result of a scan to an ansatz in
 \Ref~\cite{Hisano:2010re}. Likewise, condition
 (\ref{eq:newer_numeric_stau_condition}) also comes from a fit of the results of
 a scan to an ansatz in \Ref~\cite{Kitahara:2013lfa}. The form of
 the condition  which we present here comes from combining the definition of
 $\tan {\beta}_{\text{eff}}$ from \Eq{4} in \Ref~\cite{Kitahara:2013lfa} with
 their condition (10).

We also mention here that the dividing line between passing and failing
 condition~(\ref{eq:older_numeric_stau_condition}) is plotted in pink in the
 following figures, and condition~(\ref{eq:newer_numeric_stau_condition}) in
 grey. Again, if these lines would only appear in white regions of our figures,
 we do not plot them.

Note that conditions (\ref{eq:older_numeric_stau_condition}) and
 (\ref{eq:newer_numeric_stau_condition}) were obtained assuming that
 $A_{\tau} = 0$. Thus in the context of the CMSSM, where one needs large,
 negative $A_0$ to explain the Higgs mass by the stop loops, it can already be
 expected that these limits will not be very helpful.

We are going to discuss CMSSM scenarios where light staus and stops appear and
 point out that these scenarios are often ruled out by an unstable DSB vacuum
 which is not indicated by any of the above thumb rules.

\section{Parameter point selection and stability evaluation}
\label{sec:method}

\subsection{Evaluating the SUSY-scale Lagrangian}
Since parameter points of the CMSSM are defined by mixtures of GUT- and
 SUSY-scale parameters, it is impossible to investigate a parameter point
 without having calculated the full renormalization group running of the full
 Lagrangian from one scale to the other. Several programs are available for
 public use that do this job: 
 {\tt ISAJET} \cite{Baer:1993ae,Paige:2003mg}, 
 \softsusy\ \cite{Allanach:2001kg},
 \spheno\ \cite{Porod:2003um,Porod:2011nf}, 
 or {\tt SuSpect} \cite{Djouadi:2002ze}, for example. All of these
 codes have been shown to agree quite well \cite{Allanach:2003jw}. 
 We chose to evaluate all our points with a \spheno\ module created by 
 \sarah\ \cite{Staub:2008uz,Staub:2009bi,Staub:2010jh,Staub:2012pb} and 
 show a comparison to \softsusy in \Sec~\ref{sec:trajectory}.

On a technical level, the input flags were set such that the renormalization
 group equations (RGEs) were evaluated with two-loop beta functions. 
The following values were used for the SM parameters:
 \begin{eqnarray}
 \label{eq:SMinputs}
& \alpha^{-1}(M_Z) = 127.93,\, \ G_F = 1.166370\cdot10^{-5} \text{GeV}^{-2}
 , \,\ \alpha_S = 0.1187, \,\,
m_Z = 91.1887~\text{GeV}, & \nonumber \\
&  m_b(m_b) = 4.18~\text{GeV}, \,\, 
m_t = 173.5~\text{GeV},\,\,
m_\tau = 1.7769~\text{GeV} &
 \end{eqnarray}
 The entire mass spectrum was evaluated at one loop, \ie the known two-loop
 corrections
 \cite{Degrassi:2001yf, Dedes:2002dy, Brignole:2001jy, Brignole:2002bz,
 Degrassi:2010ne} were neglected, in order to have a consistent input to
 evaluate the one-loop effective potential without having to worry about
 whether higher orders may affect the consistency of the calculation.

\subsection{Evaluating the stability of the effective potential}
\label{sec:evaluating_stability}

The purpose of this work is to determine how stable the DSB local minimum of the
 MSSM potential is against possible deeper minima with different patterns of
 broken symmetries. Throughout our investigation we use \vcs\ 
 (version~{\tt 1.0.11}) to determine the stability of CMSSM parameter points. We
 use the full one-loop effective potential, and we refer the reader to the \vcs
 manual \cite{Camargo-Molina:2013qva} for details.

We use the MSSM model files provided by default, which were generated with
 \sarah 4 \cite{Staub:2013tta}. Unless otherwise stated, we allow six scalar
 fields to obtain non-zero \vevs, which are assumed to be real: the neutral
 components of the two Higgs doublets $H_{d}$ and $H_{u}$ (with \vevs\ \vd{} and
 \vu{} respectively), ``left-handed'' stau \sL (with \vev\ \vl{}),
 ``right-handed'' stau \sE (with \vev\ \ve{}), and one color of ``left-handed''
 and ``right-handed'' stops \sQ and \sU (with \vevs\ \vq{} and \vt{}
 respectively).

There is a certain amount of arbitrariness in deciding how long a tunneling time
 from the DSB vacuum to a different vacuum is acceptable. We chose to
 distinguish between parameter points where the tunneling time out of the DSB
 vacuum is greater than or less than $0.217$ times the age of the known Universe
 (which we take to be $13.8$ billion years), \ie three billion years. We denote
 metastable parameter points with tunneling times less than  three gigayears as
 {\it short-lived} and those with longer tunneling times as {\it long-lived}. A
 na{\"{\i}}ve estimate then of the DSB false vacuum surviving the observed
 $13.8$ gigayears with a tunneling time of three gigayears, assuming a Poisson
 distribution, is  one per-cent.  The default threshold used by \vcs if not
 overridden is $1.38$ billion years (\ie a tenth of the age of the known
 Universe), but since the tunneling time drops very rapidly (exponentially in
 the bounce action), the shift in the dividing line between a tunneling time of
 $1.38$ gigayears and three gigayears is not even visible on our plots.

In addition, we note that the tunneling time calculation depends on
 numerical routines that try to find the optimal tunneling path between the DSB
 and CCB minima which minimizes the bounce action. It might happen sometimes
 that \ct fails to find the optimal tunneling path and short-lived minima get
 labeled as long-lived. This manifests as apparently ragged borders between
 areas of short- and long-lived metastable points, and occasional islands of
 long-lived parameter points appearing in predominantly short-lived parameter
 point regions. To be conservative, we chose to keep the long-lived label for
 such points although we are quite certain that they are short-lived.

We note that there is a slight inconsistency in allowing CMSSM parameter points
 with flavor violation occuring in the Lagrangian and using model files allowing
 for only one generation of sfermions to have non-zero \vevs, as the tadpole
 equations for \vevs for the other two generations will not be satisfied in
 general for non-zero \vevs for both Higgs and third-generation sfermions.
 However, since the off-diagonal elements of the Yukawa matrices and trilinear
 soft SUSY-breaking terms are in general very small, the induced \vevs would be
 also in general very small, and evaluating the stability of the DSB minimum
 with respect to tunneling to a field configuration very close to the proper
 three-generation minimum should be sufficient. We note that this approximation
 (that small \vevs for first and second generation sfermions are neglected)
 certainly cannot produce an incorrect diagnosis of a parameter point as being
 metastable, as a three-generation minimum cannot be less deep than the nearby
 field configuration with zero \vevs for the first and second generation
 sfermions.

The applied procedure is quite conservative and we want to emphasize that any
 parameter points that we find to be metastable are definitely not absolutely
 stable, as we have found at least one counterexample minimum to the statement
 ``this parameter point does not have any deeper minima than the DSB minimum''
 regardless of whether there are even deeper minima that we have not found
 (either because of the slight increase in depth by allowing more generations to
 have non-zero \vevs or because there are deeper minima that develop at the
 one-loop level that are not found by the \vcs algorithm). Conversely, we do not
 claim that parameter points that we label as stable are guaranteed to be
 stable: we merely use it as a shorthand for the statement that we did not find
 any deeper minima and so to the best of our knowledge, the parameter point has
 a stable DSB minimum. Likewise, points found to be long-lived possibly might be
 short-lived, but short-lived points definitely are short-lived, as any deeper
 minima will not make the DSB minimum more stable against tunneling to the CCB
 minimum that we found.

The MSSM model files were validated as reproducing the correct one-loop tadpole
 equations in the vicinity of the DSB minimum, as well as the correct tree-level
 running masses, which are the important part of the one-loop corrections.
 Furthermore, the case of allowing stau \vevs (but not stop \vevs) agrees with
 the expressions of \Ref~\cite{Ferreira:2000hg}, noting the different sign
 convention for the Yukawa terms.
 
\subsection{Dependence on which scalars are allowed non-zero \vevs}
\label{sec:which_vevs}

\begin{figure}[tbp]
\centering
\includegraphics[width=0.32\linewidth
]{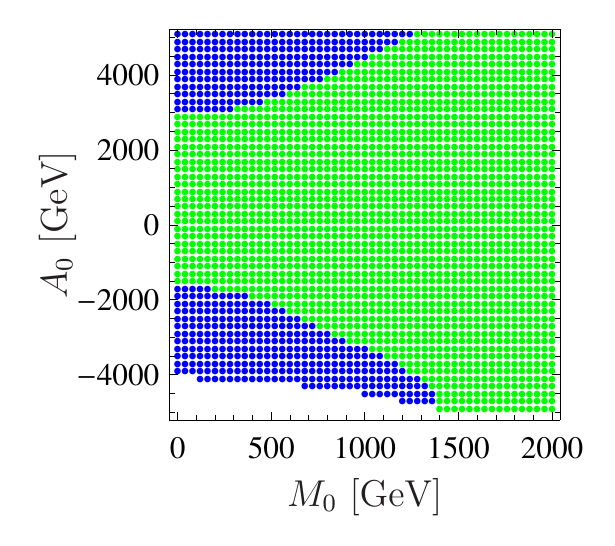} \hfill
\includegraphics[width=0.32\linewidth
]{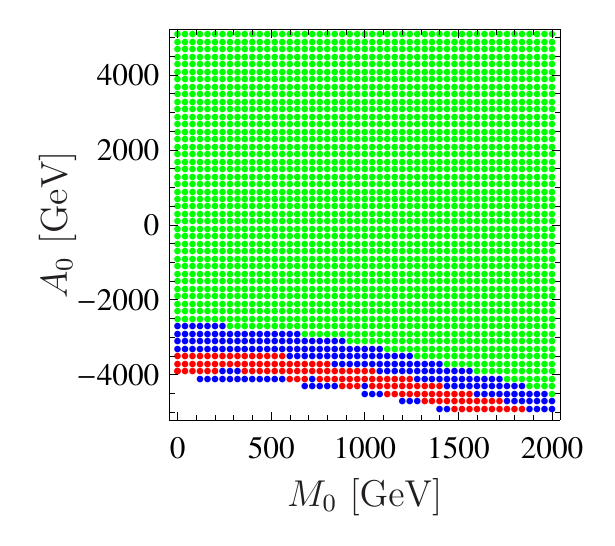} \hfill 
\includegraphics[width=0.32\linewidth
]{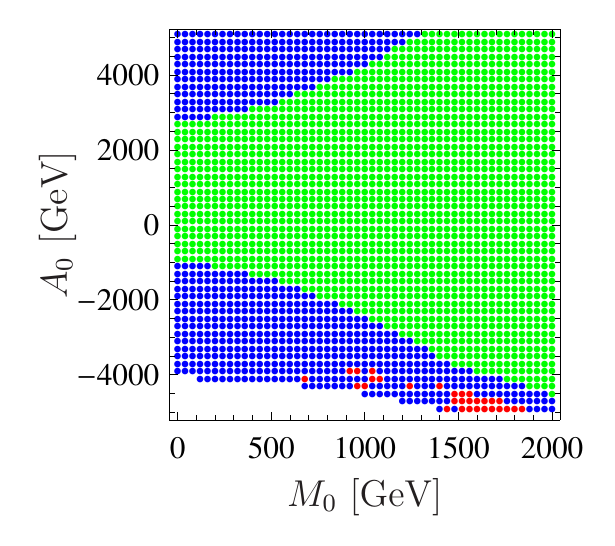}
\caption{Vacuum stability in the $(M_0,A_0)$-plane for input values of
 $\tan\beta=10$, $M_{1/2}=1$~TeV and $\mu>0$. (Here and in the other figures,
 $\tan\beta$ is taken to mean the input parameter for the DSB vacuum.) On the
 left, we allow only for stau \vevs, in the middle only for stop \vevs and on
 the right for both stau and stop \vevs. Green indicates that no CCB minimum
 deeper than the DSB minimum was found, while blue and red indicate that the DSB
 minimum is only metastable, as there is at least one deeper CCB minimum. The
 red points are short-lived, while the blue points are long-lived, compared to a
 threshold of three gigayears. The lack of a smooth boundary between red and
 blue is a numerical artifact and is discussed in the text.}
 \label{fig:stopstau}
 \end{figure}

Most checks in literature for CCB minima, which we are going to discuss in more
 detail in the following, assume that there are either stau or stop \vevs, but
 not both non-zero at the same time. However, we point out that it is not
 necessarily possible to treat them separately. In \Fig~\ref{fig:stopstau} we
 show the long- and short-lived points in the $(M_0, A_0)$ plane allowing either
 only stau or only stop or stau and stop \vevs. One can see that the trivial
 overlaying of the two special cases does not reproduce the more general case.
 The entire range of $A_0$ between -2 and +3~TeV seems to have a stable vacuum
 if one studies only stau or stop \vevs. However, allowing for both \vevs one
 can see that the vacuum is just metastable for $A_0 < -1.5$~TeV and small
 values of $M_0$. Therefore in the CMSSM it is especially necessary for
 intermediate values of $\tan\beta$ to check carefully for all possible
 combinations for stau and stop \vevs. Just in the limit of large or small
 $\tan\beta$ checks for pure stau respectively stop \vev scenarios might be
 sufficient. 

\subsection{Robustness of numerical results}
\label{sec:trajectory}

One might worry about how sensitive the presence of CCB minima is to the exact
 values used, and whether the small (understood) differences between values of
 parameters in different renormalization schemes could change the conclusion as
 to whether a parameter point is stable or metastable.

In lieu of an exhaustive duplication of plots, we compare the results along a
 line in the CMSSM parameter space using two different MSSM spectrum generators,
 with two different renormalization schemes: \spheno and \softsusy. Both are
 known to agree well on the physical observables, but the Lagrangian parameters
 differ, as \softsusy uses the modified \DRbar\ scheme described in
 \cite{Martin:2002} while we use \spheno in a scheme where the values of \vd{}
 and \vu{} at the DSB minimum do not change with loop corrections, as described
 in the appendix of \cite{Hirsch:2012kv}.

The line that we choose is that which is obtained by starting at SPS4
 \cite{Allanach:2002nj} and decreasing $A_{0}$ from zero to the point where the
 lightest neutralino is no longer the lightest supersymmetric partner. This
 entire line has been ruled out by the LHC already, as the masses of the colored
 sparticles are too light all along it, but it serves as an example.

\begin{figure}[tbp]
\centering
\includegraphics[width=0.8\linewidth
]{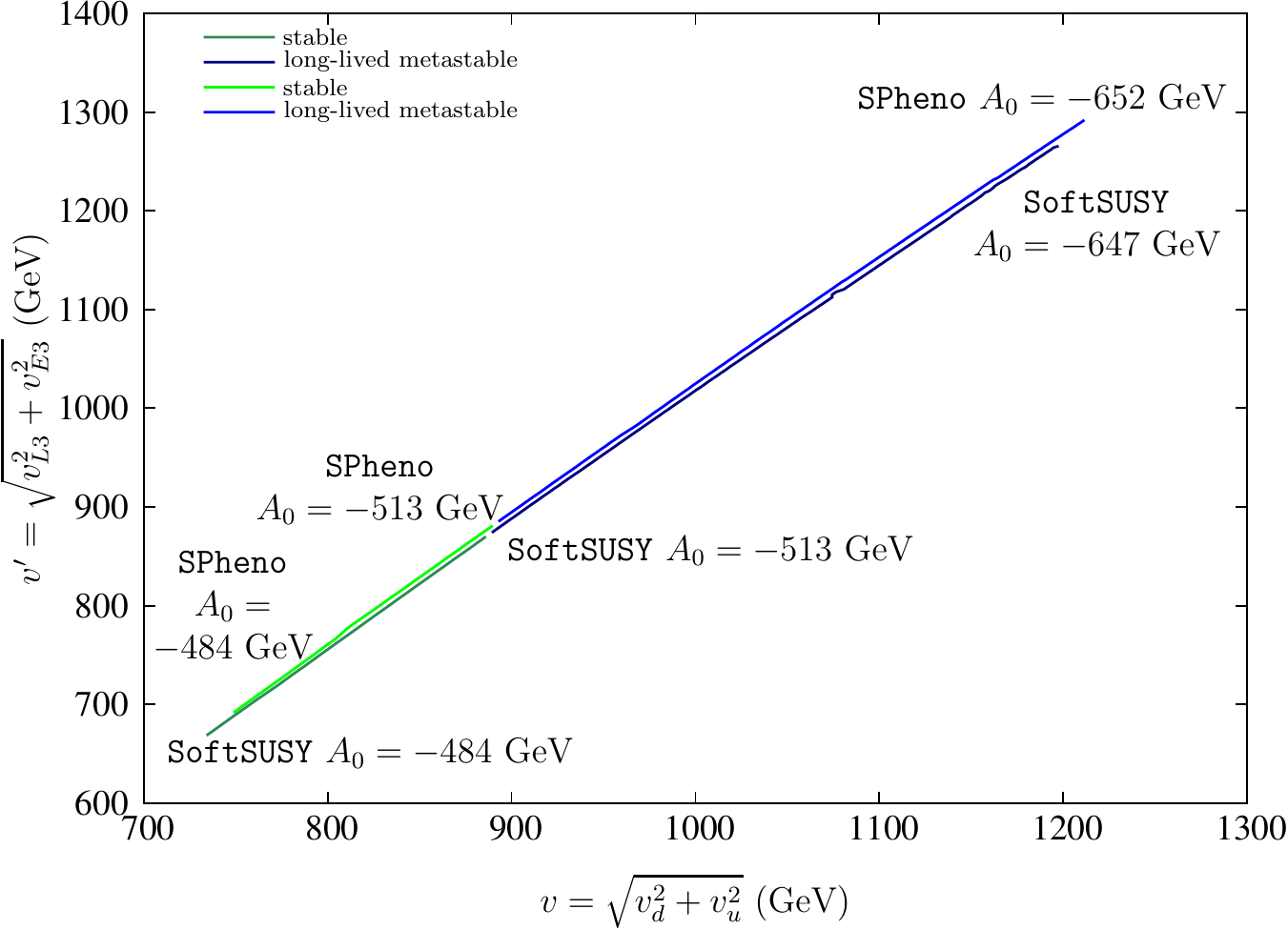}
\label{fig:trajectory_comparison}
\caption{Comparison of the evolution of a charge-breaking minimum evaluated
 with \spheno and with \softsusy, starting from SPS4 and decreasing $A_{0}$, as
 a trajectory in Higgs \vev vector length $v$ and stau \vev vector length $v'$.
 Similarly to as in \Fig~\ref{fig:stopstau}, the green segment denotes parameter
 points where the DSB vacuum is stable, and the blue where it is long-lived
 metastable. Both sets of line segments are very close together. The line
 evaluated with \softsusy has darker shades for its segments than the line
 evaluated with \spheno has.}
\end{figure}

In \Fig~\ref{fig:trajectory_comparison}, we show a projection of the
 co-ordinates of the DSB mininum of SPS4 and the first other
 minimum of the potential (excluding minima related to the input minimum by
 gauge transformations) that develops, in the plane of the length
 $v = \sqrt{ v_{d}^{2} + v_{u}^{2} }$ of the Higgs \vevs and the length
 $v' = \sqrt{ v_{L3}^{2} + v_{E3}^{2} }$ of the stau \vevs. (As in
 \Sec~\ref{sec:rules}, the \vev of ${\tilde{\tau}}_{L}$ is
 given by $\vl{}/\sqrt{2}$ and that of ${\tilde{\tau}}_{R}$ by
 $\ve{}/\sqrt{2}$.) The inputs of SPS4
 ($M_{0} = 400 \gev, M_{1/2} = 300 \gev, \tan \beta = 50, |\mu| > 0$) were held
 constant while $A_{0}$ was varied from $0 \gev$ (the input for SPS4) to the
 value where even the DSB minimum is phenomenologically uninteresting as the
 stau is the LSP ($-652 \gev$ according to \spheno, $-647 \gev$ according to
 \softsusy). The input parameters were run with \spheno and with \softsusy from
 the GUT scale to the SUSY
 scale~$\sqrt{ m_{{\tilde{t}}_{1}} m_{{\tilde{t}}_{2}} }$.

The CCB minimum that develops along this line in parameter space only comes into
 existence once $|A_{0}|$ gets large enough, and as $|A_{0}|$ increases, it
 becomes deeper and eventually becomes deeper than the DSB minimum, as one might
 have expected. Both \spheno and \softsusy agreed that the CCB minimum developed
 at $A_{0} = -431 \gev$ and that it became deeper than the DSB minimum at
 $A_{0} = -513 \gev$. The details of the tunneling time for larger values of
 $|A_{0}|$ did not match exactly but were nevertheless acceptably close, given
 that the tunneling time varies exponentially in the bounce action.

Both generators also yielded values for the \vevs at the CCB minima that agreed
 within the ranges expected for the differences between the programs, as can be
 seen in \Fig~\ref{fig:trajectory_comparison} and in
 \Tab~\ref{tab:trajectory_vevs}. Furthermore, as one can see in
 \Tab~\ref{tab:trajectory_vevs}, the values of the \vevs at the CCB minimum do
 not vary wildly as $A_{0}$ is varied, and they remain quite safely within a few
 times the renormalization scale; hence the loop expansion should be
 trustworthy. For completeness, we note that, as expected, the \vevs increase for
 larger $|A_{0}|$ and appear to be tending towards ratios that minimize the
 $D$-term contribution to the effective potential. Furthermore, one can clearly
 see how much the ratios of \vd{}, \vl{}, and \ve{} at the global minima deviate
 from the ratio 1:1:1 that is assumed in the derivation of
 condition~(\ref{eq:useless_stau_condition}). Finally, we note in passing that
 conditions~({\ref{eq:useless_stau_condition}})
 --~(\ref{eq:newer_numeric_stau_condition}) are all satisfied along the entire
 line in parameter space from $A_{0} = 0 \gev$ to $A_{0} = -652 \gev$.

\begin{table}
\centering
\begin{tabular}{
c@{\hspace{4mm}}c@{\hspace{4mm}}c@{\hspace{4mm}}c@{\hspace{4mm}}c@{\hspace{4mm}}c}
 $A_{0}$ & generator & \vd{} & \vu{} & \vl{} & \ve{} \\
\hline
 -484 & \spheno & 184 & 726 & 409 & 558 \\
 -484 & \softsusy & 181 & 712 & 394 & 540 \\
\hline
 -513 & \spheno & 269 & 851 & 540 & 701 \\
 -513 & \softsusy & 274 & 846 & 532 & 694 \\
\hline
 -652 & \spheno & 485 & 1110 & 819 & 999 \\
 -647 & \softsusy & 481 & 1097 & 800 & 981 \\
\hline
\end{tabular}
\caption{Full \vev configurations of the CCB minimum that develops by changing
 the input of SPS4 by making $A_{0}$ negative, for endpoints of the segments of
 \Fig~\ref{fig:trajectory_comparison}, where all dimensionful quantities are to
 be understood as in units of GeV.}
\label{tab:trajectory_vevs}
\end{table}

\subsection{Scale and loop order dependence}

\begin{figure}[tbp]
\centering
\includegraphics[width=0.4\linewidth
]{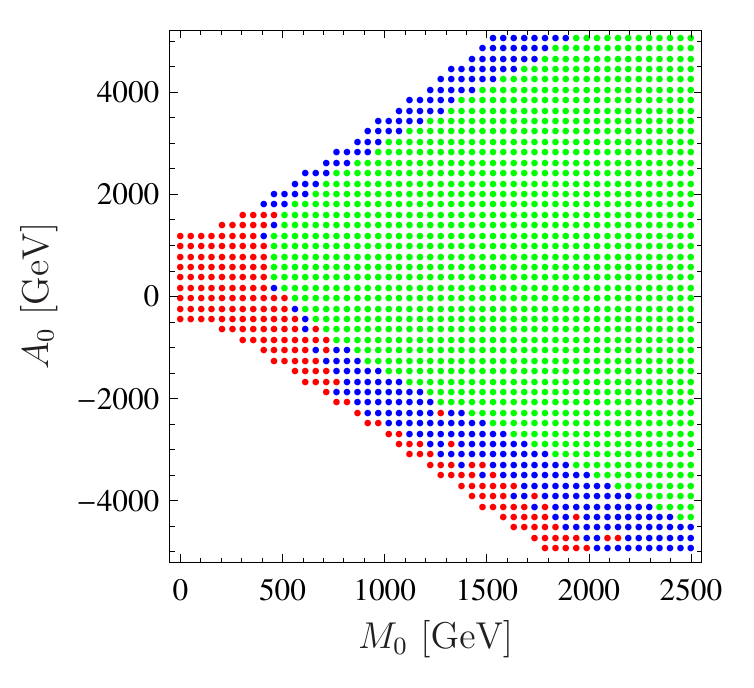} \hfill
\includegraphics[width=0.4\linewidth
]{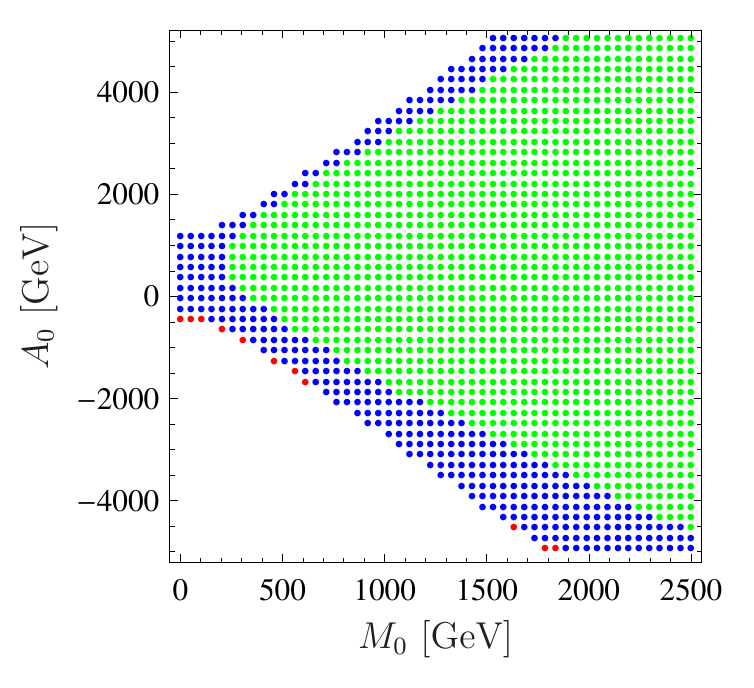} \vspace{-0.4cm}\\
\includegraphics[width=0.4\linewidth
]{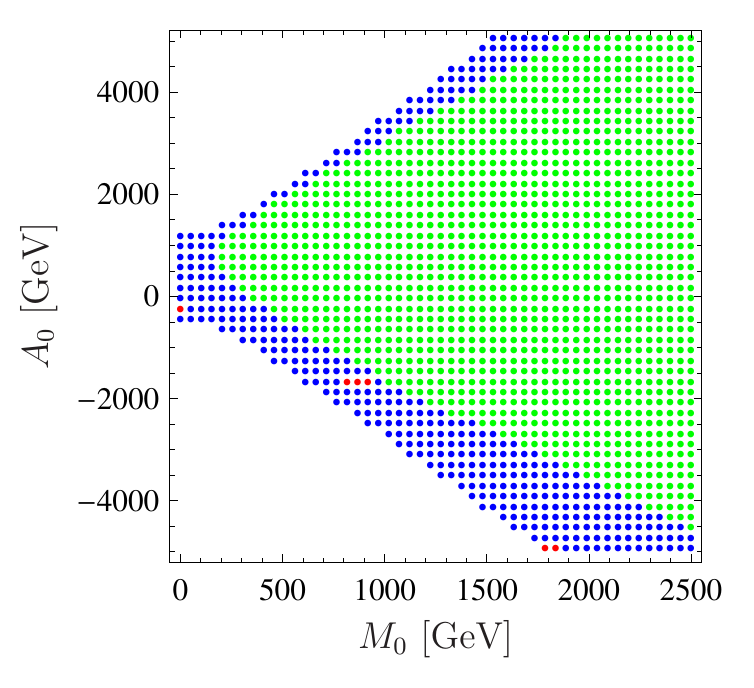} \hfill
\includegraphics[width=0.4\linewidth
]{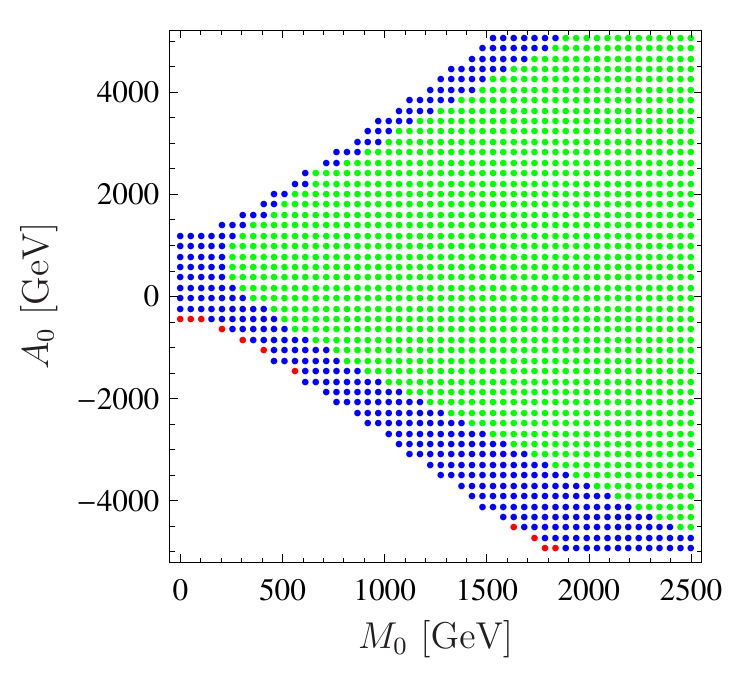}  \vspace{-0.4cm} \\
\includegraphics[width=0.4\linewidth
]{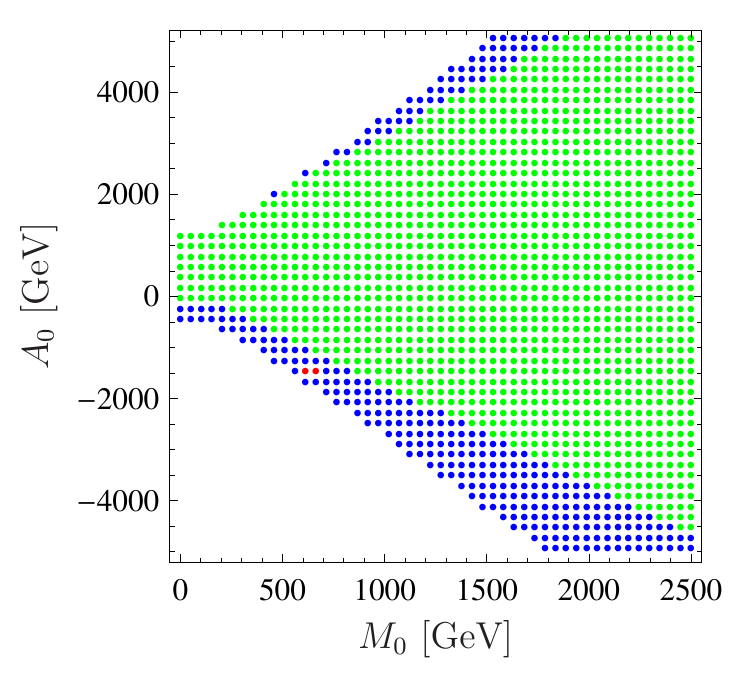} \hfill
\includegraphics[width=0.4\linewidth
]{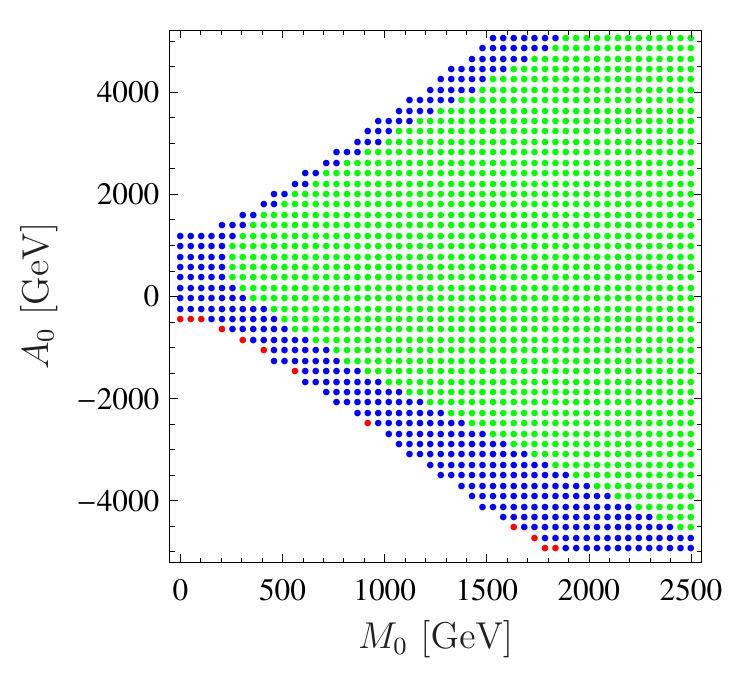} 
\caption{Vacuum stability in the $(M_0, A_0)$ plane for fixed
 $M_{1/2} = 1$~TeV, $\mu>0$ and $\tan\beta=40$. On the left, only the tree-level
 potential was considered. On the right, the full one-loop effective potential
 was taken into account. The renormalization scale was
$Q = \sqrt{ m_{{\tilde{t}}_{1}} m_{{\tilde{t}}_{2}} } / 2$ (first row),
 $Q = \sqrt{ m_{{\tilde{t}}_{1}} m_{{\tilde{t}}_{2}} }$ (second row), and
 $Q = 2 \sqrt{ m_{{\tilde{t}}_{1}} m_{{\tilde{t}}_{2}} }$ (last row). 
 The color code is the same as in \Fig~\ref{fig:stopstau}.}
\label{fig:Q_dependence}
\end{figure}

It is well known that parameters like masses or cross-sections suffer usually
 from a sizable scale dependence when they are calculated just at tree level. To
 reduce this dependence on the renormalization scale $Q$, higher-order
 corrections have to be taken into account. In the MSSM, the geometric average
 $\sqrt{ m_{{\tilde{t}}_{1}} m_{{\tilde{t}}_{2}} }$ of the stop masses is
 usually taken as renormalization scale, since it is expected that the $Q$
 dependence of the Higgs mass is minimized arround this scale. A similar
 statement can be made about the vacuum stability: if one just considers the
 tree-level potential, the areas with CCB minima can be significantly shifted by
 changing the renormalization scale. This is shown in the left column of
 \Fig~\ref{fig:Q_dependence}, where we show the results for
 $Q = \sqrt{ m_{{\tilde{t}}_{1}} m_{{\tilde{t}}_{2}} } / 2$,
 $Q = \sqrt{ m_{{\tilde{t}}_{1}} m_{{\tilde{t}}_{2}} }$, and
 $Q = 2 \sqrt{ m_{{\tilde{t}}_{1}} m_{{\tilde{t}}_{2}} }$. However, if one goes
 to the full one-loop effective potential, the sensitivity on the scale is
 significantly reduced, as depicted in the right column of
 \Fig~\ref{fig:Q_dependence}. For completeness we note that that even for
 $A_0=0$ one still has sizable $A$-parameters proportional to $-M_{1/2}$ 
 at  the electroweak scale which cause the instability of the potential for
 small $M_0$.

We also note that this is an explicit example that undermines arguments that
 radiative effects do not change tree-level conclusions on the absolute
 stability of vacua such as in \cite{PhysRevD.48.4352}, as it is plain to see
 that even relatively small changes of scale can radically change tree-level
 conclusions.

The interesting possibility of bound states composed of stops has been
 suggested in the literature for extremely large $|A_{t}|$ \cite{Pearce:2013yja}
 and the effective Lagrangian in terms of these bound states could have yet more
 minima which could be deeper than the DSB minimum. However, such a mechanism is
 very non-perturbative and does not look possible within the CMSSM, at least in
 the parameter regions which we investigate in this paper, and also certainly
 the existence of extra minima to which the DSB false vacuum can tunnel cannot
 {\it reduce} the decay width hence cannot increase the tunneling time.

\subsection{Thermal effects}

Of course, even if a parameter point has an acceptably long tunneling time out
 of the DSB vacuum at zero temperature, it may be the case that the false vacuum
 would not survive a period at a sufficiently high temperature below the
 critical temperature where the undesired minimum becomes less deep. While the
 one-loop thermal corrections are known \cite{Brignole:1993wv} and implemented
 in \ct \cite{Wainwright:2011kj}, the evaluation of the minimal bounce action at
 non-zero temperature is currently unfeasibly slow for our work, and conclusions
 on the stability of a parameter point are strongly dependent on the assumed
 thermal history of the Universe. Hence we postpone such an investigation for
 future work.

\section{Constraining relevant regions of the CMSSM parameter space}
\label{sec:constraining}

\subsection{Constraining $A_0$ and $\tan\beta$}

\begin{figure}[tbp]
\centering
\includegraphics[width=0.45\linewidth
]{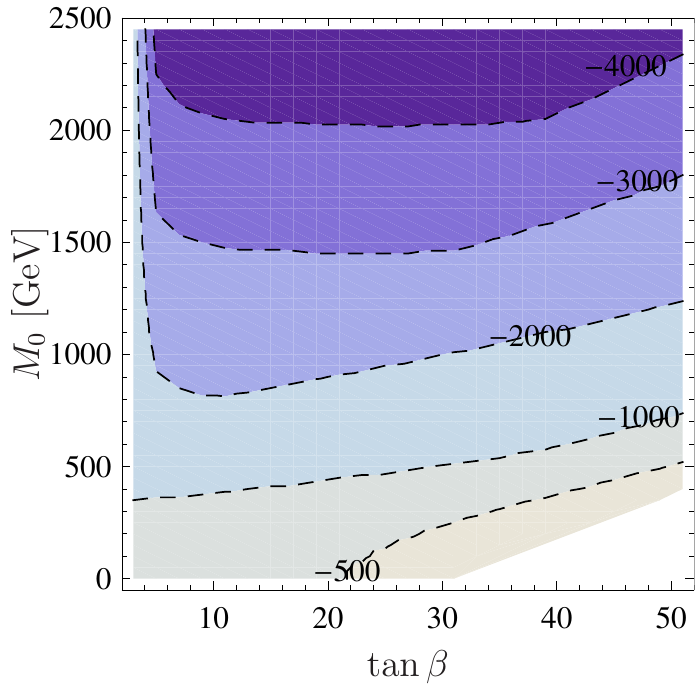} \hfill
\includegraphics[width=0.45\linewidth
]{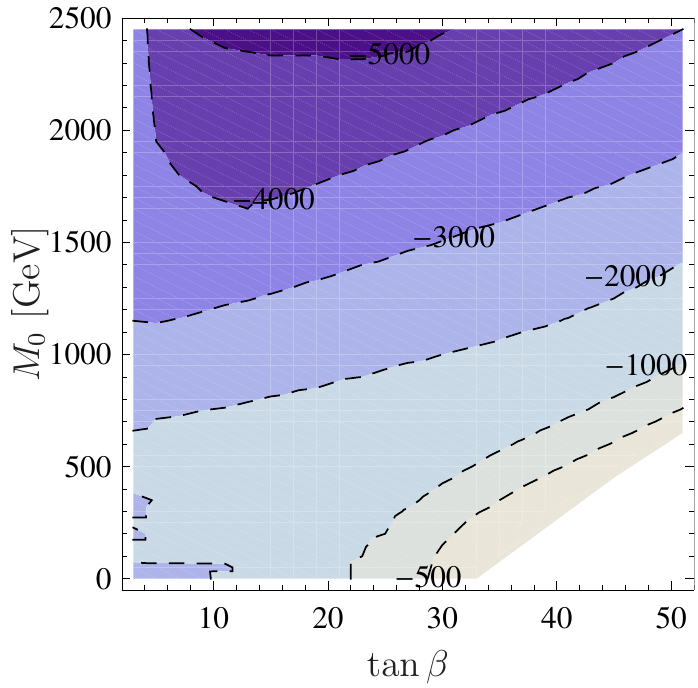} \\
\includegraphics[width=0.45\linewidth
]{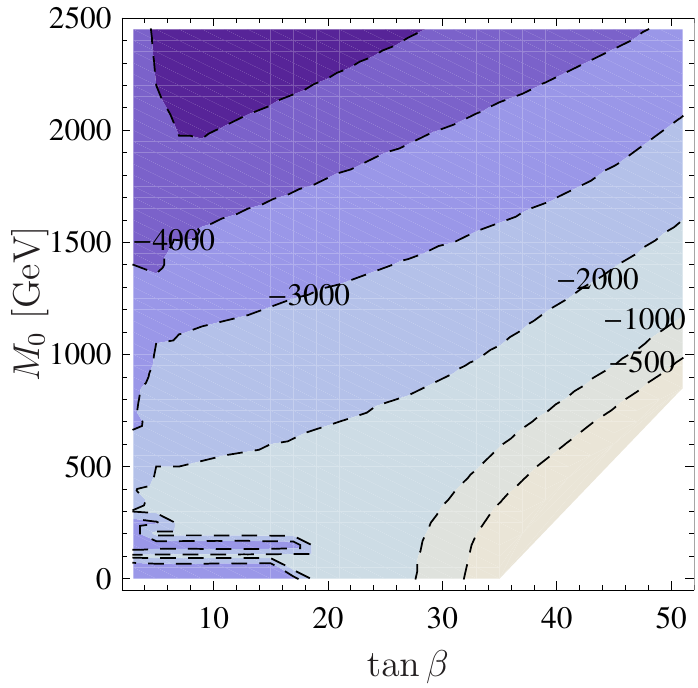} \hfill
\includegraphics[width=0.45\linewidth
]{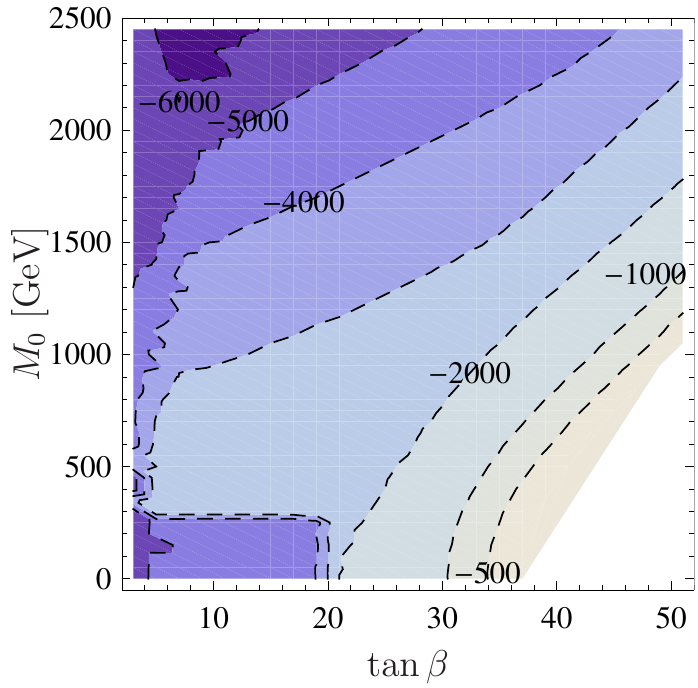}
\caption{Minimal value allowed for $A_0$ in the $(\tan\beta, M_0)$ plane to have
 a stable DSB vacuum. We used $\mu>0$ and $M_{1/2} = 500$~GeV (upper left),
 $M_{1/2} = 1000$~GeV (upper right), $M_{1/2} = 1500$~GeV (lower left) and
 $M_{1/2} = 2000$~GeV (lower right).}
\label{fig:minA0}
\end{figure}

We start with a check for regions in the CMSSM space which don't have any CCB
 vacuum deeper than the  DSB minimum. This gives us already some hint as to
 which regions of the parameter space are `safe', and also where one has to
 perform more dedicated checks, as discussed in the following. For this purpose
 we look for the maximal value of $|A_{0}|$ allowed for a given combination of
 $M_0$, $M_{1/2}$, $\tan\beta$ and sign($\mu$), assuming that $A_{0} < 0$. We
 concentrate on negative $A_0$ as it has some preference due to the Higgs mass
 measurements \cite{Arbey:2011ab}. The result is shown in \Fig~\ref{fig:minA0}. 
 Note that this also restricts the smallest mass allowed for the lighter stau
 and stop, as discussed in sections \ref{sec:staus} and \ref{sec:stops}. We see
 that the maximal value of $|A_{0}|$ allowed is rather restricted, especially
 for large $\tan\beta$. The main reason is that the larger $\tan\beta$ gets,
 the smaller $m^{2}_{{\tilde{\tau}}_{R}}$ gets while keeping all other
 parameters fixed. That the electroweak minimum is destabilized for large
 $\tan\beta$ is a feature which we are going to see often in the following.

\begin{figure}[tbp]
\centering
\includegraphics[width=0.8\linewidth]{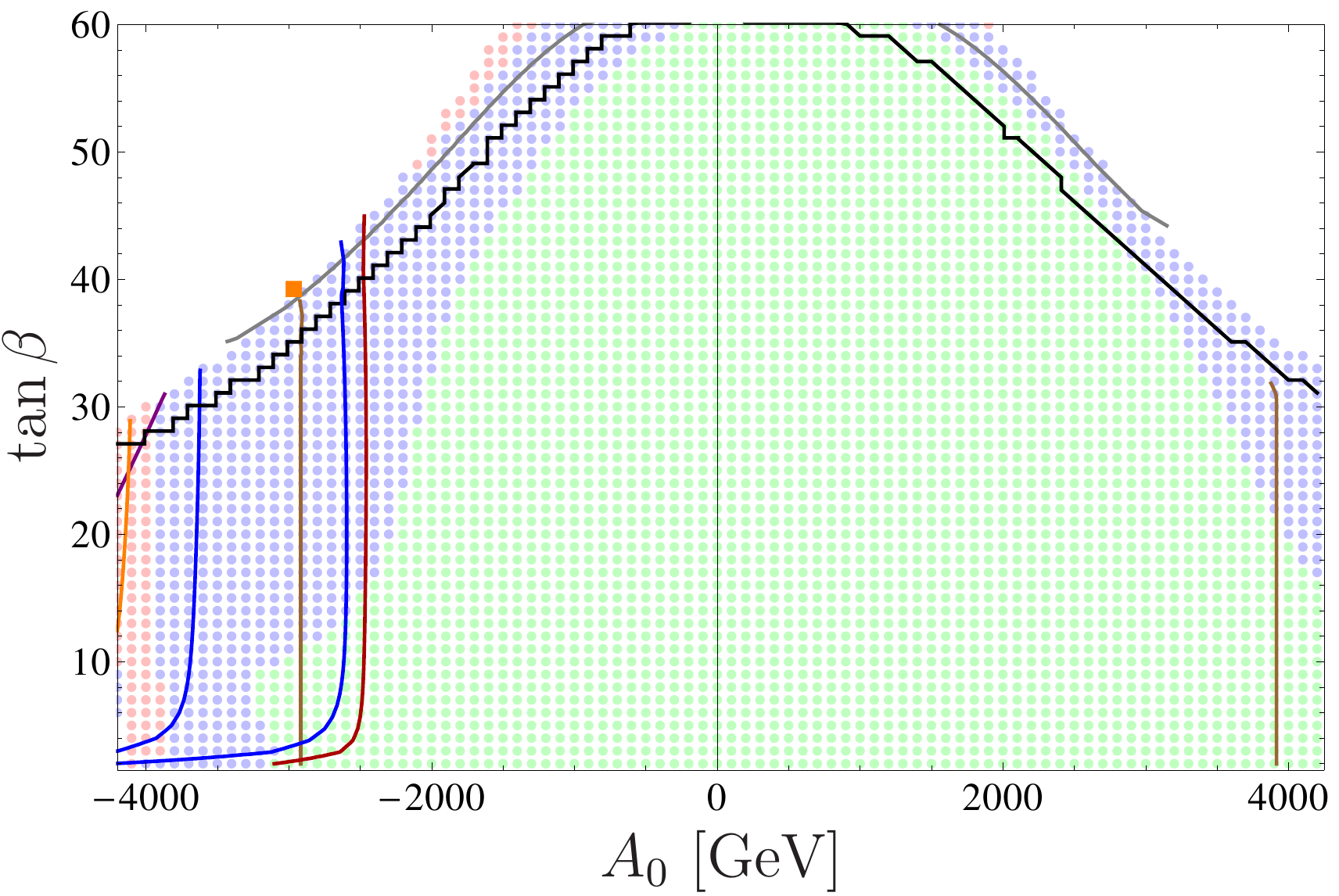}
\caption{Vacuum stability in the $(A_0, \tan\beta)$-plane for fixed values of
 $M_0 = M_{1/2} = 1$~TeV and $\mu > 0$. The color coding is as in
 \Fig~\ref{fig:stopstau} and in \Sec~\ref{sec:rules}: points on the other side
 of the solid lines from the $A_{0} = 0$ axis fail the corresponding conditions
 and so would be identified as having CCB minima deeper than the DSB minima. The
 left-most of the blue lines corresponds to taking $0.85^{2}$ in
 condition~(\ref{eq:stop_large_tb_condition}), and the other to $0.65^{2}$. The
 single orange square corresponds to a projection of the best-fit point of
 \Ref~\cite{Bechtle:2012zk} for reference. Points below the dotted line have the
 lightest neutralino as the LSP.}
\label{fig:A0_against_tanbeta}
\end{figure}

As a next step, we investigate the problem of deeper minima further and check
 not only if there are deeper vacua but also if the DSB vacuum would be long- or
 short-lived. As mentioned in \Sec~\ref{sec:evaluating_stability}, we take a
 relatively conservative threshold of $0.217$ times the observed life time of
 the known Universe (corresponding to a one per-cent survival probability) to
 categorize metastable points as long-lived or short-lived. We present the
 distribution of stable, long- and short-lived vacua according to this
 definition in the $(\tan\beta, A_0)$ plane for fixed values of
 $M_0 = M_{1/2} = 1$~TeV and $\mu > 0$ in \Fig~\ref{fig:A0_against_tanbeta}. We
 show also a comparison with the regions excluded by the limits using the
 conditions~(\ref{eq:useless_GUT_condition})
 to~(\ref{eq:newer_numeric_stau_condition}). One can see that at least some
 analytical limits produce the qualitative dependence on $\tan\beta$ but not
 one of them is really accurate and they miss many short-lived vacua. Note that
 those limits do not distinguish between long- and short-lived vacua but only
 assess the presence of a deeper CCB vacuum. Hence one has to compare them with 
 the division between the stable (green) areas and the metastable (blue and red)
 areas, rather than how well they would separate long-lived (blue) from
 short-lived (red). It is then clear that even a combination of all of those
 rules would not exclude about half the points of our scans with CCB vacua
 deeper than the DSB vacua. 

We note that the values of $M_{1/2}$ and $M_{0}$ of $1$ TeV are rather close to
 those of the central values of CMSSM best-fit point after including a mass for
 the Higgs boson of $126 \gev$ of \cite{Bechtle:2012zk}
 ($M_{1/2} = 1167.4{}^{+594.0}_{-513.0} \gev,
 M_{0} = 1163.2{}^{+1185.3}_{-985.7} \gev, \tan \beta = 39.3{}^{+16.7}_{-32.7},
 A_{0} = -2969.1{}^{+6297.8}_{-1234.9}$), and that the central values for
 $\tan \beta$ and $A_{0}$ are within the long-lived metastable region of
 \Fig~\ref{fig:A0_against_tanbeta}. This seems to be rather endemic to models
 where the stau mass and the stop mass are related, for trying to fit both the
 dark matter relic abundance, requiring light staus, and a relatively heavy
 Higgs boson, requiring heavy stops, thus pushing the fit to large trilinear
 terms in the CMSSM. It also occurs regardless of allowing non-universal Higgs
 masses: the central values of the NUHM1 ``low'' best-fit point of
 \cite{Buchmueller:2012hv} also result in a CCB global vacuum. Therefore, we
 continue with a more detailed study of the light stau parameter space.

\subsection{Constraining the light stau parameter space}
\label{sec:staus}

\begin{figure}
\centering
\includegraphics[width=0.45\linewidth
]{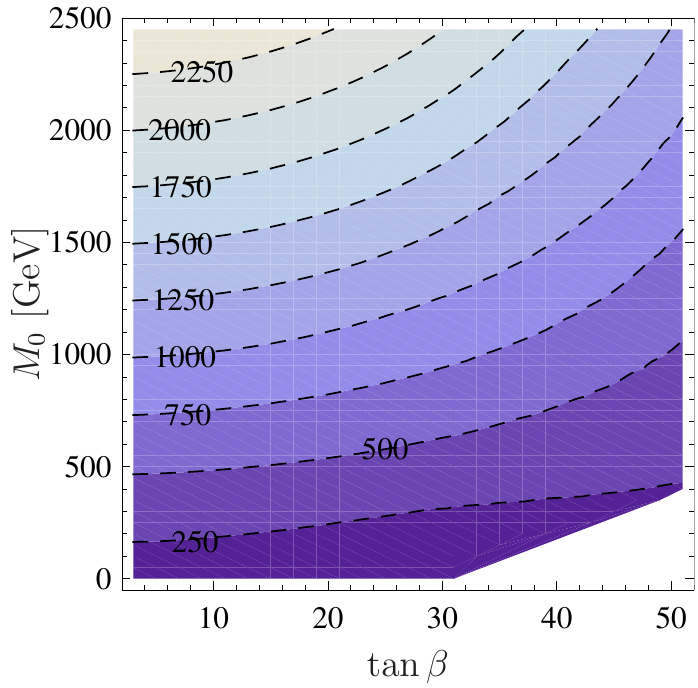} \hfill
\includegraphics[width=0.45\linewidth
]{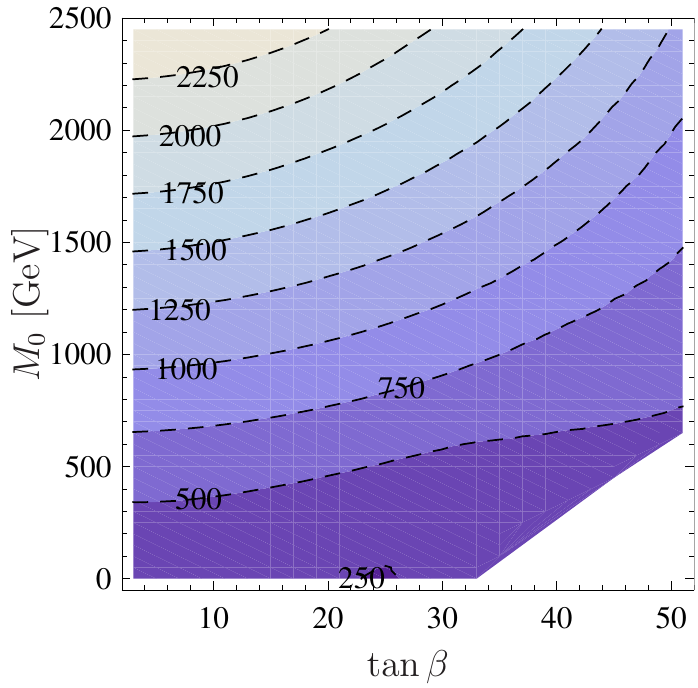} \\
\includegraphics[width=0.45\linewidth
]{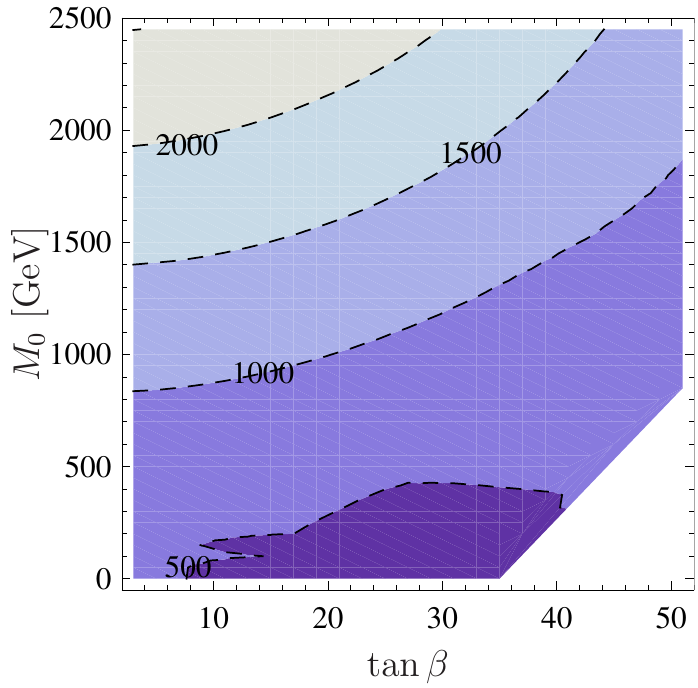} \hfill
\includegraphics[width=0.45\linewidth
]{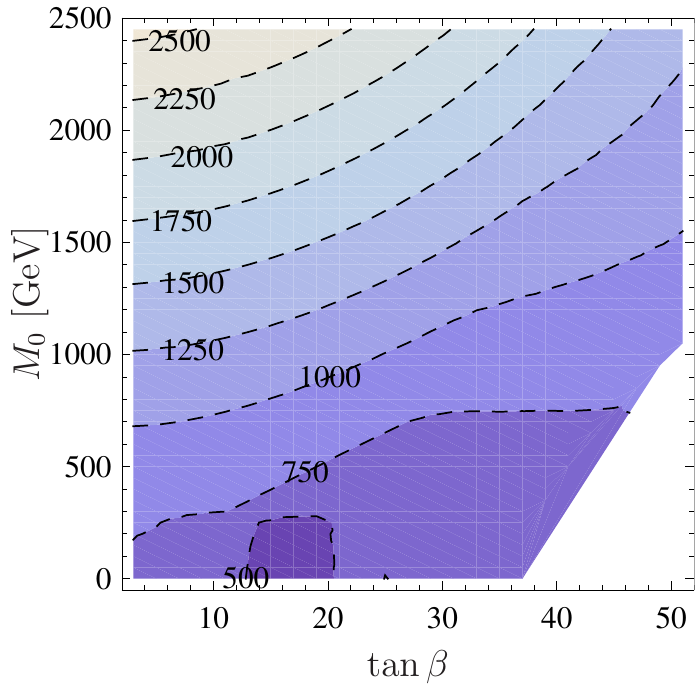}
\caption{Minimal stau mass in the $(\tan\beta, M_0)$ plane to have a stable
 DSB vacuum for $A_0<0$. We used $\mu>0$ and $M_{1/2} = 500$~GeV (upper
 left), $M_{1/2} = 1000$~GeV (upper right), $M_{1/2} = 1500$~GeV (lower left)
 and $M_{1/2} = 2000$~GeV (lower right).}
\label{fig:minStau}
\end{figure}

Light staus in the MSSM are particularly interesting because if their mass is
 sufficiently close to the mass of a neutralino LSP then co-annihilation
 between the staus and the neutralinos can explain the observed dark matter
 relic density \cite{Ellis:2001zk, Ellis:2003cw, Ellis:2012aa}. The measured
 dark matter relic density in the $3 \sigma$ range of Planck \cite{Ade:2013ktc}
 is $\Omega h^2= 0.1199\pm 0.081$. This demands a mass splitting of at most a
 few GeV between the stau and neutralino. To obtain this, usually large values
 of $|A_{0}|$ and/or $\tan\beta$ are needed. However, as we have seen, this is
 exactly the parameter range which is in particular danger of CCB minima due to
 stau \vevs. As a starting point we can use the results of \Fig~\ref{fig:minA0}
 for the minimal value of $A_0$ and translate this into a lower limit on the
 light stau mass for given parameters, which we show in \Fig~\ref{fig:minStau}.
 This shows that for $M_{1/2} > 500$~GeV and stau masses below 500~GeV, deeper
 vacua are possible, and therefore a careful test of the vacuum stability is
 required.

\begin{figure}[tbp]
\centering
\includegraphics[width=0.48\linewidth
]{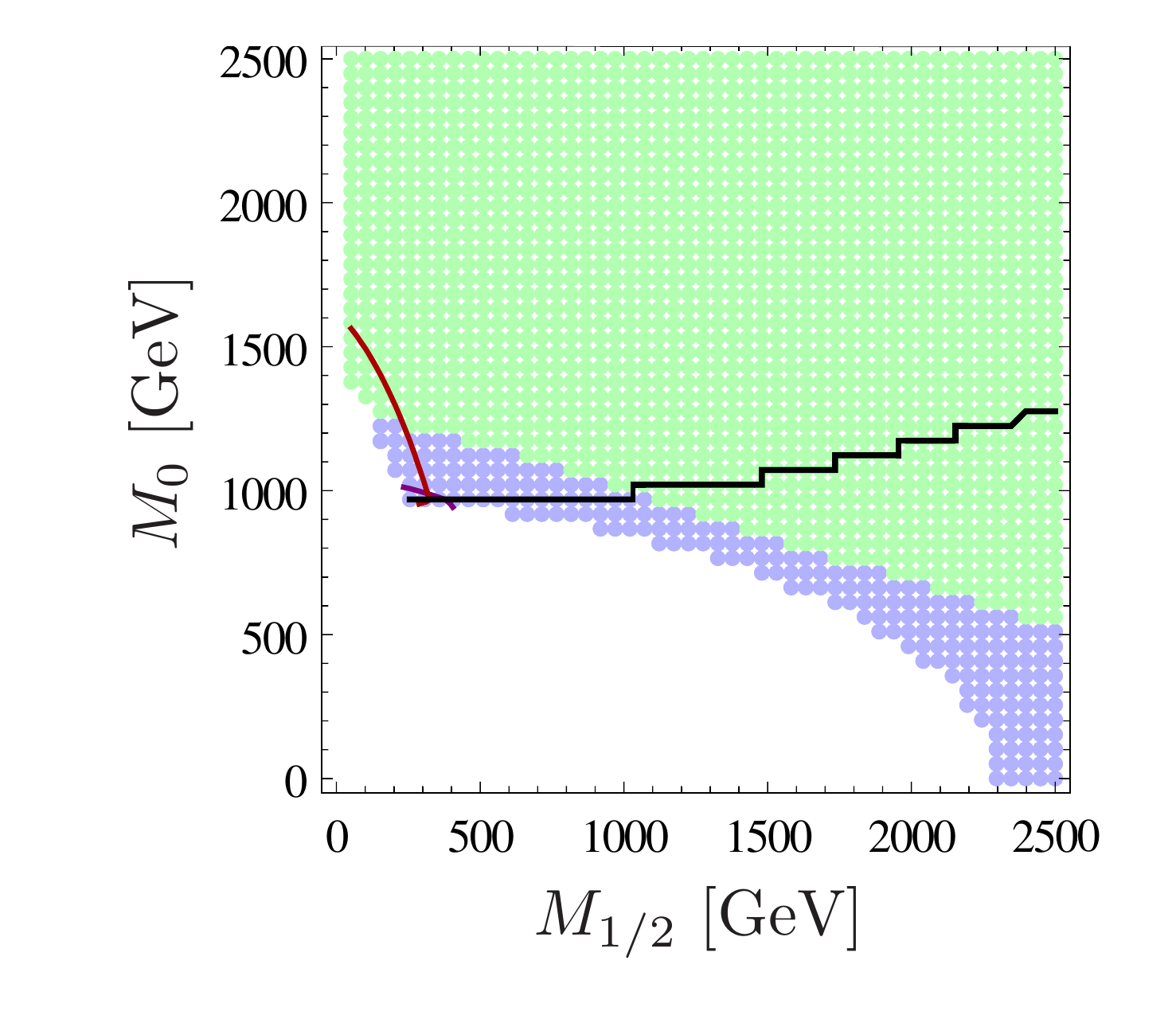} \hfill 
\includegraphics[width=0.48\linewidth
]{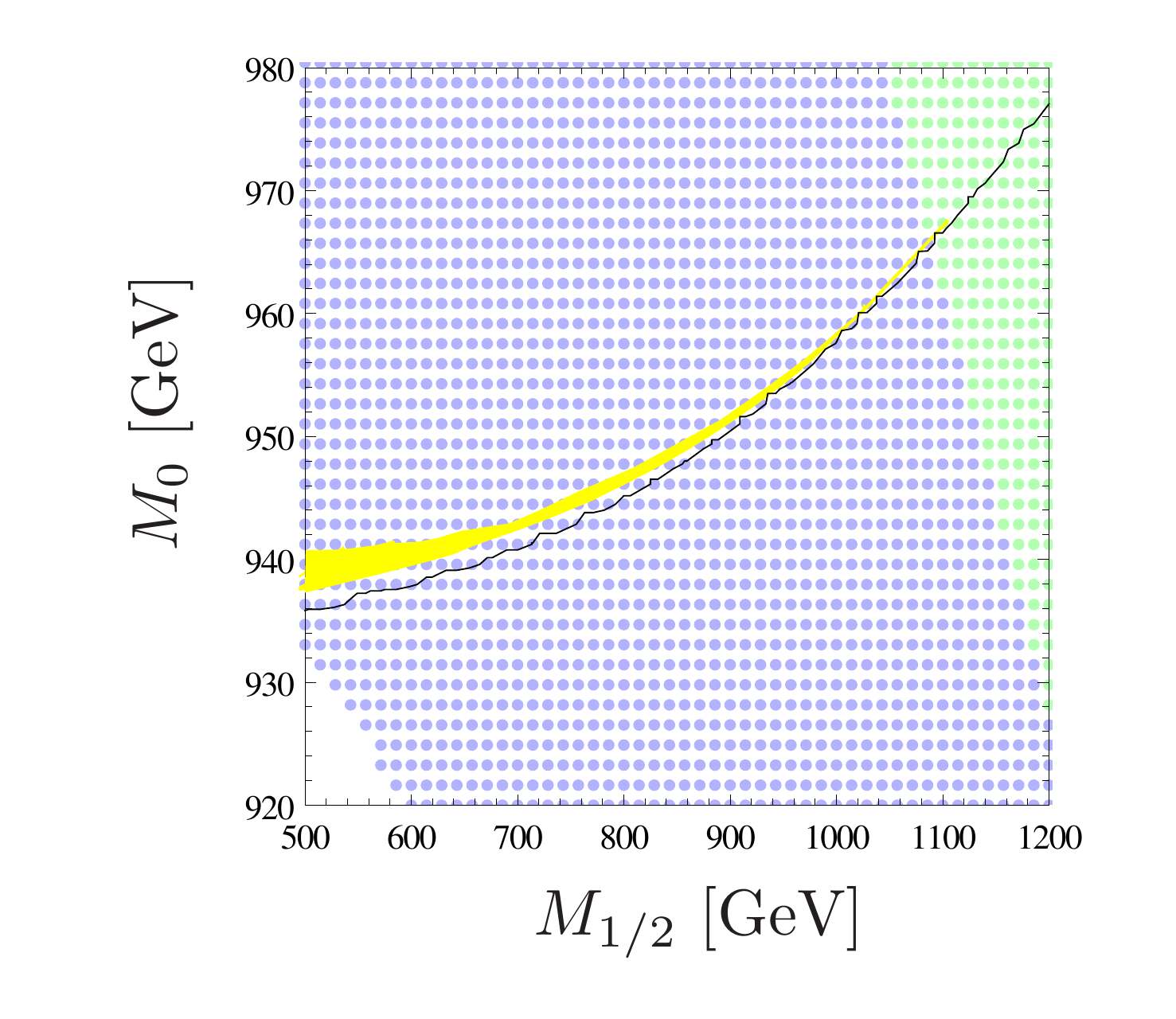} 
\caption{Dark matter and vacuum stability in the $(M_{1/2}, M_0)$ plane with
 $A_0=3$~TeV, $\mu>0$ and $\tan\beta=40$. On the left, the dashed black line
 shows the transition between a neutralino and stau LSP (stau LSP beneath the
 line); on the right we zoom in on the interesting range for dark matter: the
 yellow bands show the region where $\Omega h^2= 0.1199\pm 0.081$. The
 constraints from vacuum stability allowing for stop and stau \vevs are
 indicated. The color coding is as in \Fig~\ref{fig:stopstau} and in
 \Sec~\ref{sec:rules}: points to the left of the solid lines fail the
 corresponding conditions. Here and in subsequent figures, physical quantities
 such as the dark matter relic density and particles masses are taken to be
 evaluated at the DSB vacuum regardless of its stability.}
\label{fig:stau_coann}
\end{figure}

However, parameter points with CCB minima deeper than the DSB minimum may still
 be viable if the tunneling time is sufficiently long. We exemplify this in
 \Fig~\ref{fig:stau_coann}, where we plot the stable and metastable areas in the
 $(M_{1/2}, M_0)$ plane for $\tan\beta=40$ and $A_0=3000$~GeV. Note that this is
 the same area which has been studied in \Ref~\cite{Ellis:2012aa}. It is also
 depicted that only a very tiny region of the entire plane would be forbidden by
 the rules presented in \Sec~\ref{sec:rules} (including some stable points that
 would be mistakenly excluded), while our full numerical checks show that the
 excluded parameter region is much larger if one demands conservatively that the
 DSB minimum is the global minimum. On the right-hand side of
 \Fig~\ref{fig:stau_coann}, we zoom into the area with sufficiently large stau
 co-annihilation to explain the dark matter abundance. Here, we used
 {\tt MicrOmegas 2.4.5} \cite{Belanger:2006is,Belanger:2001fz} together with the
 {\tt SUSY Toolbox} \cite{Staub:2011dp} to calculate the relic density.
 Fortunately for advocates of the stau co-annihilation region, the tunneling
 time out of the DSB false vacuum is phenomenologically acceptably long over the
 entire region shown in \Fig~\ref{fig:stau_coann}.
 
\begin{figure}[tbp]
\centering
\includegraphics[width=0.48\linewidth]{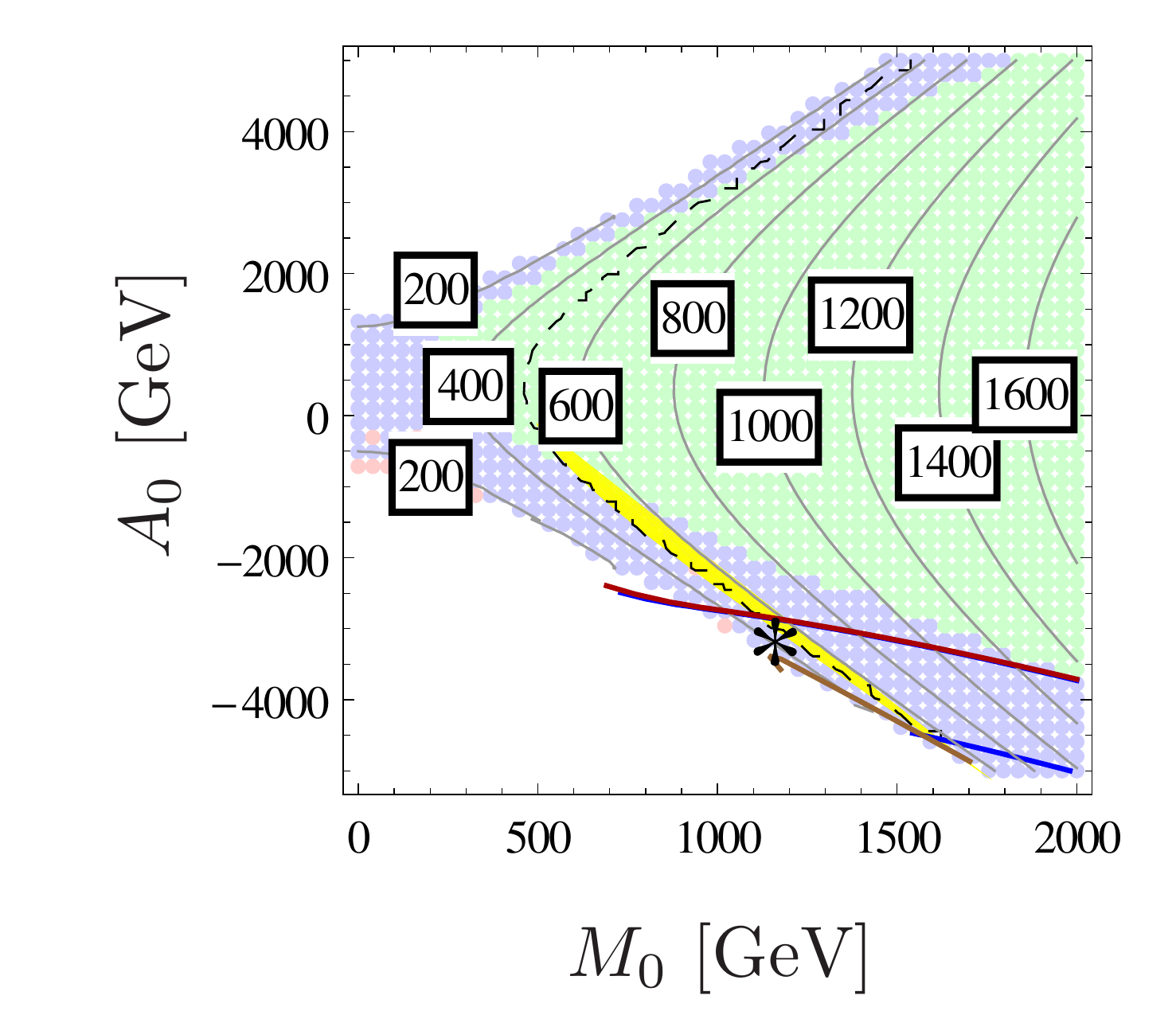}  \hfill 
\caption{Mass of the light stau and vacuum stability in the $(M_0, A_0)$ plane
 with $M_{1/2}=1167.4$~GeV, $\mu>0$ and $\tan\beta=39.3$. In the yellow region,
 the abundance of the LSP is in agreement with dark matter constraints, and the
 dashed line shows the transition to a charged LSP. The color coding is as in
 \Fig~\ref{fig:stopstau} and in \Sec~\ref{sec:rules}: points below the solid
 lines fail the corresponding conditions. The lower blue line corresponds to
 condition~(\ref{eq:stop_large_tb_condition}) taking $0.85^{2}$, the upper,
 which is almost exactly degenerate with the dark red line of
 condition~(\ref{eq:useless_sup_condition}), correponds to taking $0.65^{2}$.
 The star indicates the best-fit point according to \Ref~\cite{Bechtle:2012zk}.}
 \label{fig:fittino_DM}
\end{figure}

In \Fig~\ref{fig:fittino_DM}, we provide another example to point out that
 vacuum stability in the context of stau co-annihilation is a severe issue
 which has not been taken into account to the demanded level. Here we show the
 $(M_0, A_0)$-plane around the best-fit point found in
 \Ref~\cite{Bechtle:2012zk}. In this plane, the main part of the
 co-annihilation strip, including the best-fit point itself, would be ruled out
 by demanding that the DSB vacuum is absolutely stable, though it is allowed if
 one only demands that it has a lifetime of at least three gigayears. Again, the
 rules from \Sec~\ref{sec:rules} do not constrain the interesting region at all,
 except for condition~(\ref{eq:useless_sup_condition}), which also excludes
 stable regions. It is interesting to note that the only condition which happens
 to constrain even partially any interesting region in our scans is one that is
 being used inappropriately, as it is being applied to parameters where the
 assumptions for its derivation are invalid, and it is no surprise that it also
 excludes stable points.

\begin{figure}[tbp]
\centering
\includegraphics[width=0.45\linewidth
]{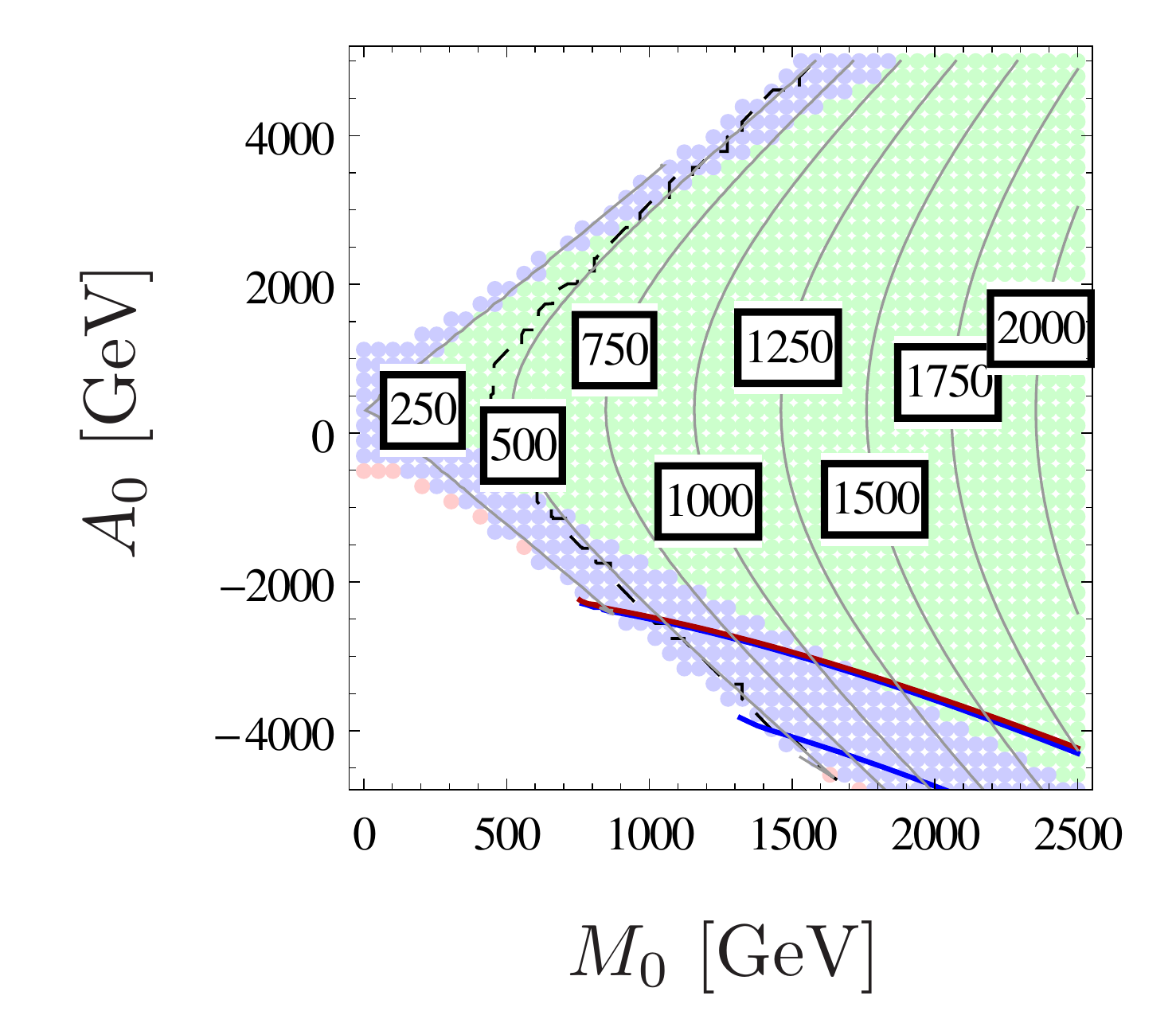}
\hfill
\includegraphics[width=0.45\linewidth
]{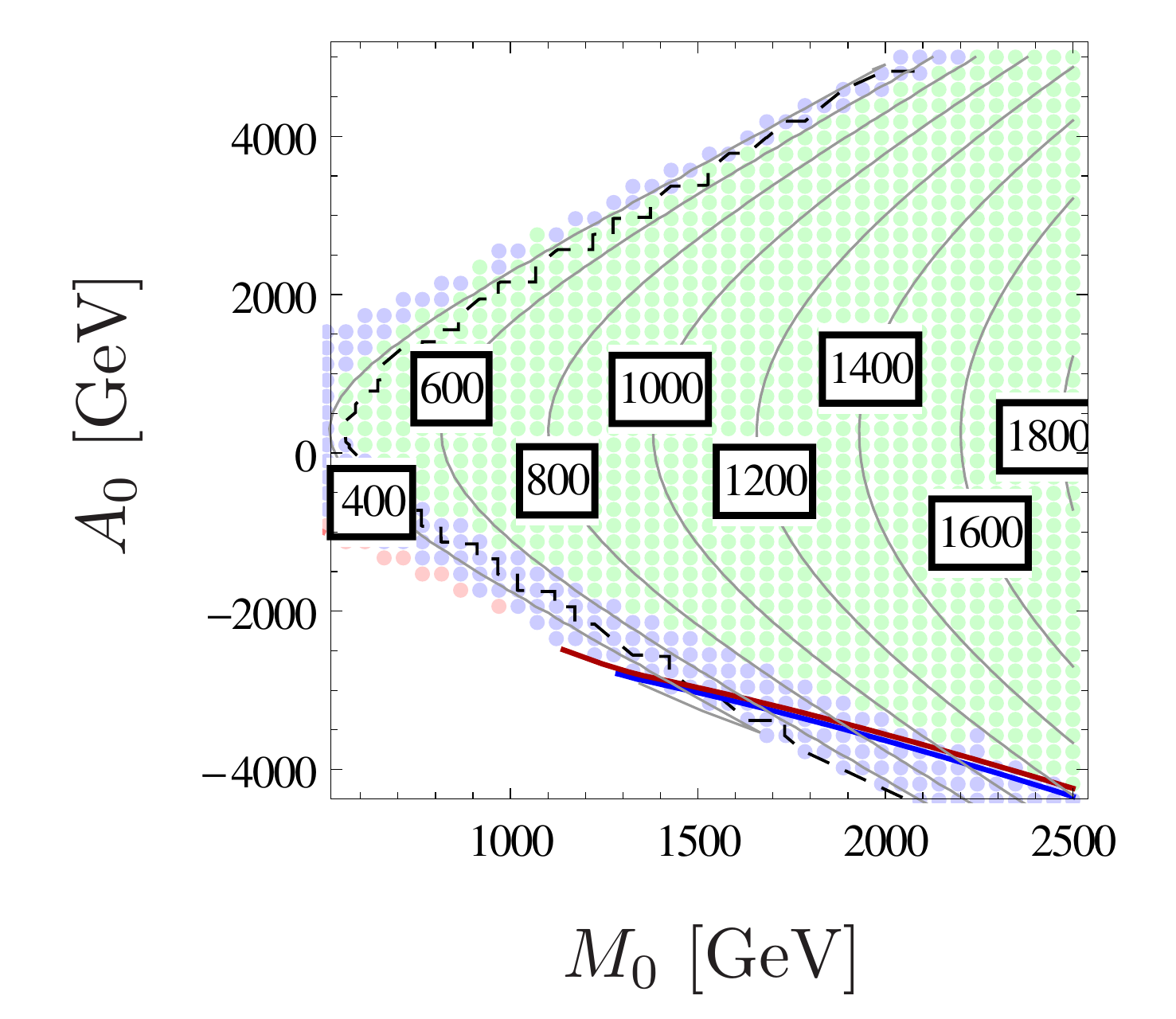} 
\caption{Vacuum stability and stau masses in the $(M_0, A_0)$ plane for fixed
 $M_{1/2} = 1$~TeV, $\mu>0$ and $\tan\beta=40$ (left) or $\tan\beta=50$ (right).
 The dashed black line shows the transition to a charged LSP (neutralino LSP to
 the right of the line). The color coding is as in \Fig~\ref{fig:stopstau} and
 in \Sec~\ref{sec:rules}: points to the left of the solid lines fail the
 corresponding conditions. As in \Fig~\ref{fig:fittino_DM}, the blue line for
 condition~(\ref{eq:stop_large_tb_condition}) with $0.65^{2}$ is almost
 degenerate with the dark red line of condition~(\ref{eq:useless_sup_condition})
 and the line for $0.85^{2}$ only excludes points with even more negative
 $A_{0}$ (and is not visible on the right).}
 \label{fig:stau_comparison}
\end{figure}

Lastly, we would like to discuss another comparison between the results of the
 full-fledged analysis and the analytical approximations of
 \Sec~\ref{sec:rules}. In \Fig~\ref{fig:stau_comparison}, we show the light
 stau masses as well as the vacuum stability in the $(M_0, A_0)$ plane for
 $\tan\beta=40$ and 50. $M_{1/2}$ was fixed to 1~TeV. For $\tan\beta=40$, the
 approximations totally fail, with the exception of
 (\ref{eq:useless_sup_condition}), which is at least not totally useless even
 though it really should only apply for $Y_{t} \ll 1$. All of the other
 conditions would lie in the white region, which is already excluded by
 tachyonic states at the DSB vacuum. For $\tan\beta=50$ the situation is only
 slightly better: tiny parts of the small $M_0$ areas are in conflict with
 conditions~(\ref{eq:older_numeric_stau_condition}). However, the main part of
 the metastable vacua would also survive all cuts applying those rules (with
 (\ref{eq:useless_sup_condition}) performing less well). Hence, for large
 $\tan\beta$ and reasonable SUSY spectra above the LHC exclusion limits, one can
 not rely at all on conditions~(\ref{eq:useless_stau_condition}),
 (\ref{eq:useless_stop_condition}), (\ref{eq:useless_GUT_condition}),
 (\ref{eq:older_numeric_stau_condition})
 or~(\ref{eq:newer_numeric_stau_condition}).

\begin{figure}[tbp]
\centering
\includegraphics[width=0.45\linewidth
]{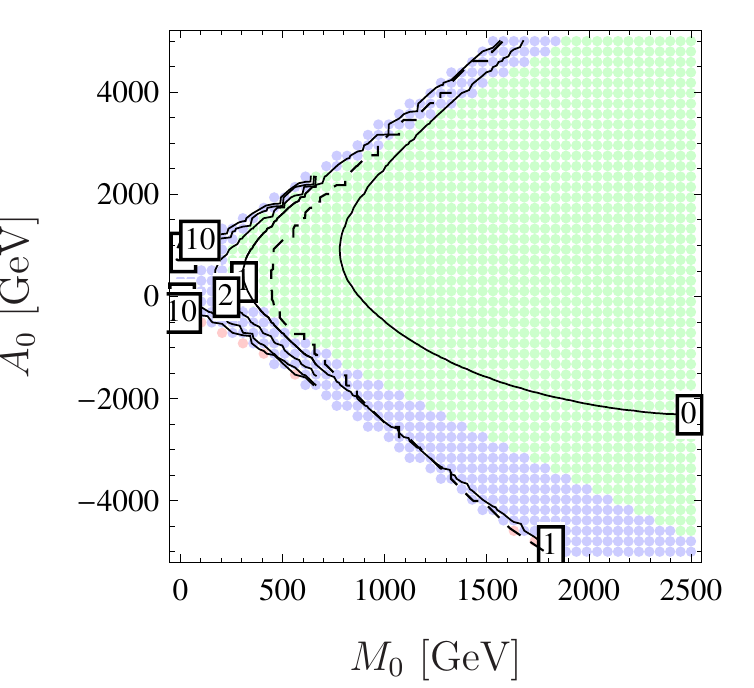}
\hfill
\includegraphics[width=0.45\linewidth
]{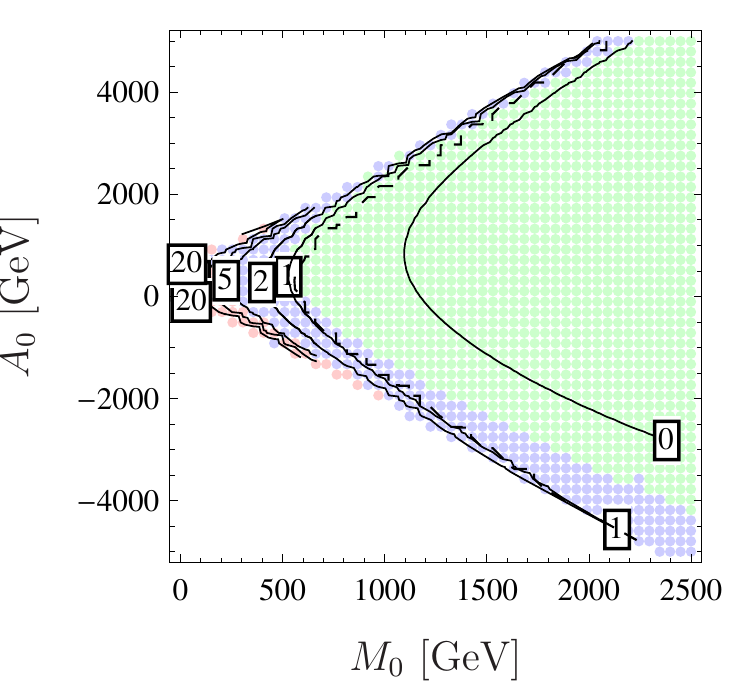}
\caption{The enhancement of the branching ratio of the lighter Higgs boson
 relative to the SM prediction, given in per-cent, in the $(M_0, A_0)$ plane
 for fixed $M_{1/2} = 1$~TeV, $\mu>0$ and $\tan\beta=40$ (left) or
 $\tan\beta=50$ (right). The contour labeled $0$ shows where the branching
 ratio is equal to that of the SM, the one labeled $1$ shows where it is $1\%$
 greater than in the SM, and so on. The dashed black line shows the transition
 to a charged LSP (neutralino LSP to the right of the line). The color coding
 is as in \Fig~\ref{fig:stopstau}.}
 \label{fig:diphoton_BRs}
\end{figure}

One might wonder whether the lower bounds on the mass of the lightest stau by
 demanding stability of the DSB vacuum could restrict parameter regions of the
 CMSSM where the branching ratio of the lighter neutral Higgs boson to a pair
 of photons is enhanced relative to the SM prediction, as is currently weakly
 favored by Atlas \cite{Aad:2013wqa}. Certainly light stau masses are required,
 close to the  LEP limit of $\mathcal{O}(100 \gev)$
 \cite{Carena:2012gp, Giudice:2012pf}, and restrictions of the pMSSM parameter
 space has been discussed to some extent in the literature
 \cite{Carena:2012mw}. We show in \Fig~\ref{fig:diphoton_BRs} the contours of
 enhancement of the ratio for two particular slices of the parameter space. As
 can be seen, enhancements of up to $20\%$ can be achieved, and that the
 tunneling time to CCB vacua is a very relevant constraint for the parameter
 region, but the regions plotted at least are not phenomenologically
 interesting anyway as the stau is much lighter than the lightest neutralino
 there, and the regions where the neutralino is a valid dark matter candidate
 also have very little enhancement of the diphoton branching ratio. This might
 favor additional light charged fermions in extended SUSY models to explain the
 possible enhancement
 \cite{SchmidtHoberg:2012yy, Benakli:2012cy, Joglekar:2013zya}, or very light 
 charginos in the MSSM below 100~GeV which could escape all present collider 
 limits \cite{Batell:2013bka}.

\subsection{Constraining the light stop parameter space}
\label{sec:stops}

The stops play an important role in the MSSM as they are needed to push the mass
 of the light Higgs to about $125$~GeV via radiative corrections
 \cite{Draper:2011aa, Heinemeyer:2011aa, Brummer:2012ns, Djouadi:2013vqa,%
 Arbey:2012bp}. For this one needs either rather heavy stops, as for example in
 gauge-mediated SUSY-breaking \cite{Ajaib:2012vc, Brummer:2012ns} or large
 $|A_t|$ to obtain the maximal mixing scenario
 \cite{Haber:1996fp, Carena:2000dp, Draper:2011aa}. A rough estimate is given by
 $A_t \simeq 0.2 A_0 - 2 M_{1/2}$ \cite{Bartl:2001wc} causing the observed
 preference for negative $A_0$. Moreover, $|A_0| \simeq 2 M_0$
 \cite{Carena:2000dp} is required implying the peril of developing CCB minima.

\begin{figure}[tbp]
\centering
\includegraphics[width=0.45\linewidth
]{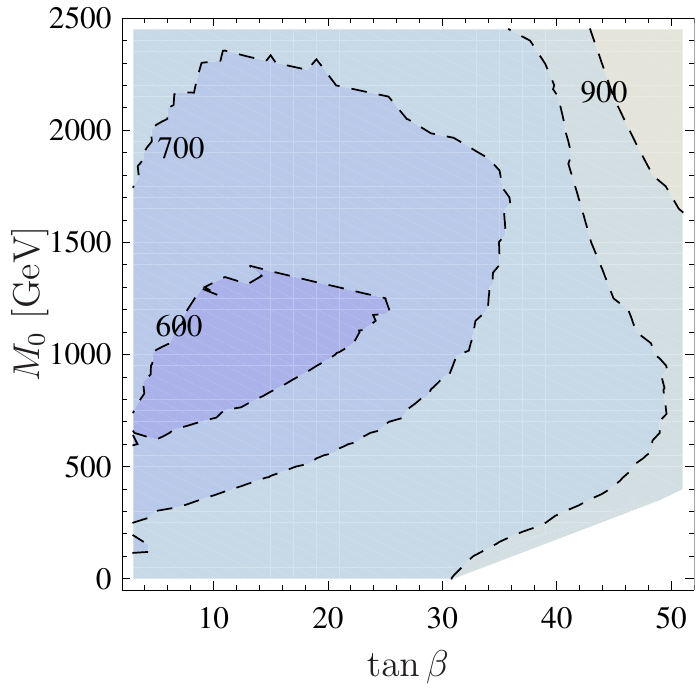} \hfill
\includegraphics[width=0.45\linewidth
]{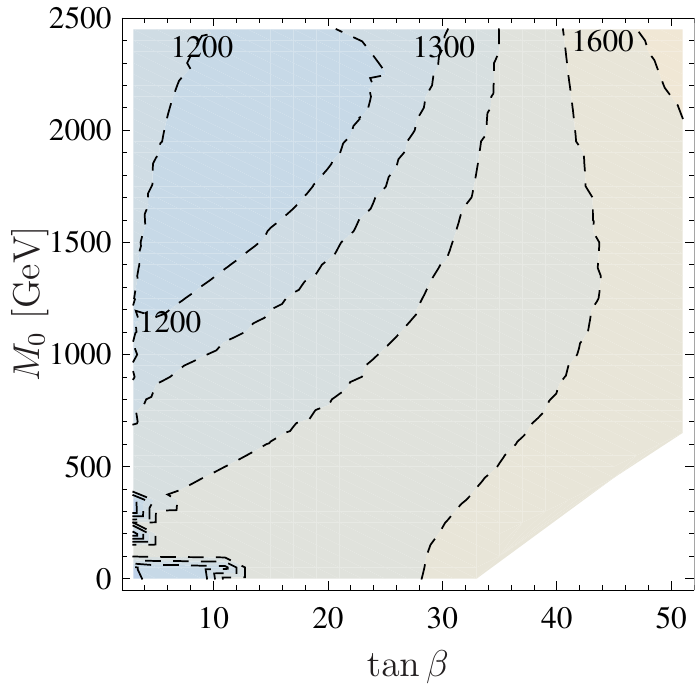} \\
\includegraphics[width=0.45\linewidth
]{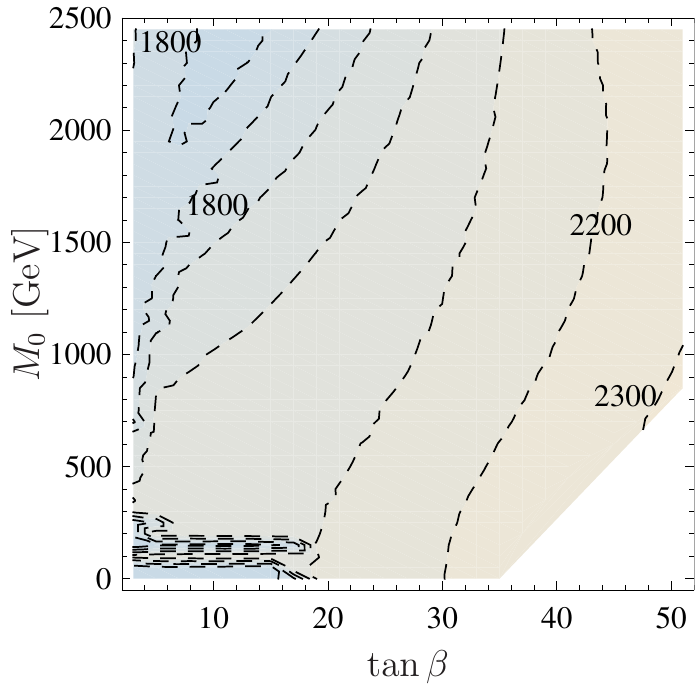} \hfill
\includegraphics[width=0.45\linewidth
]{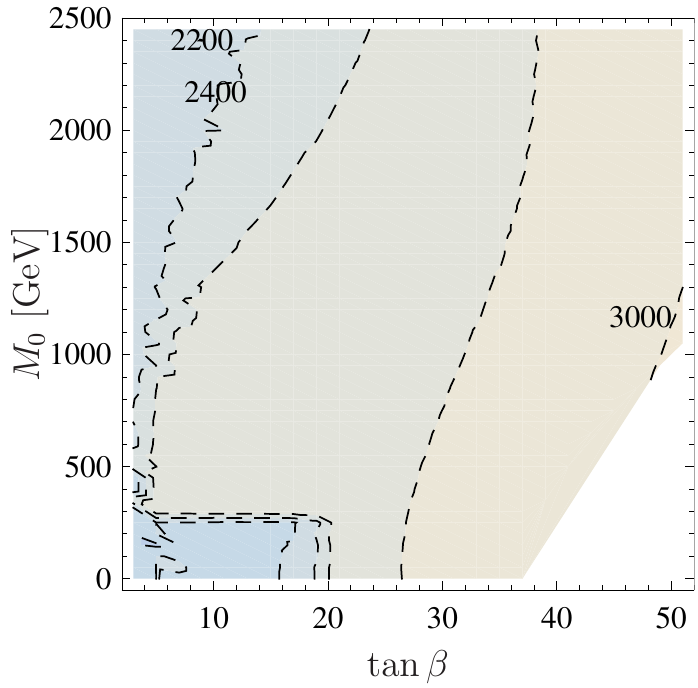}
\caption{Minimal stop mass in the $(\tan\beta, M_0)$ plane to have a stable
 DSB vacuum for $A_0<0$. We used $\mu>0$ and $M_{1/2} = 500$~GeV (upper
 left), $M_{1/2} = 1000$~GeV (upper right), $M_{1/2} = 1500$~GeV (lower left)
 and $M_{1/2} = 2000$~GeV (lower right).}
\label{fig:minStop}
\end{figure}

The information of \Fig~\ref{fig:minA0} can be translated into a lower limit on
 the stop mass by demanding a stable DSB vacuum which is given in
 \Fig~\ref{fig:minStop}. We see that for small $M_{1/2}$, this condition
 excludes light stop masses of the order of current lower limit from direct
 searches of about 600~GeV \cite{ATLAS-CONF-2013-053}. However, for larger
 values of $M_{1/2}$, the lower mass limit is much stronger: in the TeV range.
 Furthermore, our limits are independent of the mass splitting between
 the stop and the lightest neutralino or chargino.

\begin{figure}[tbp]
\centering
\includegraphics[width=0.46\linewidth]{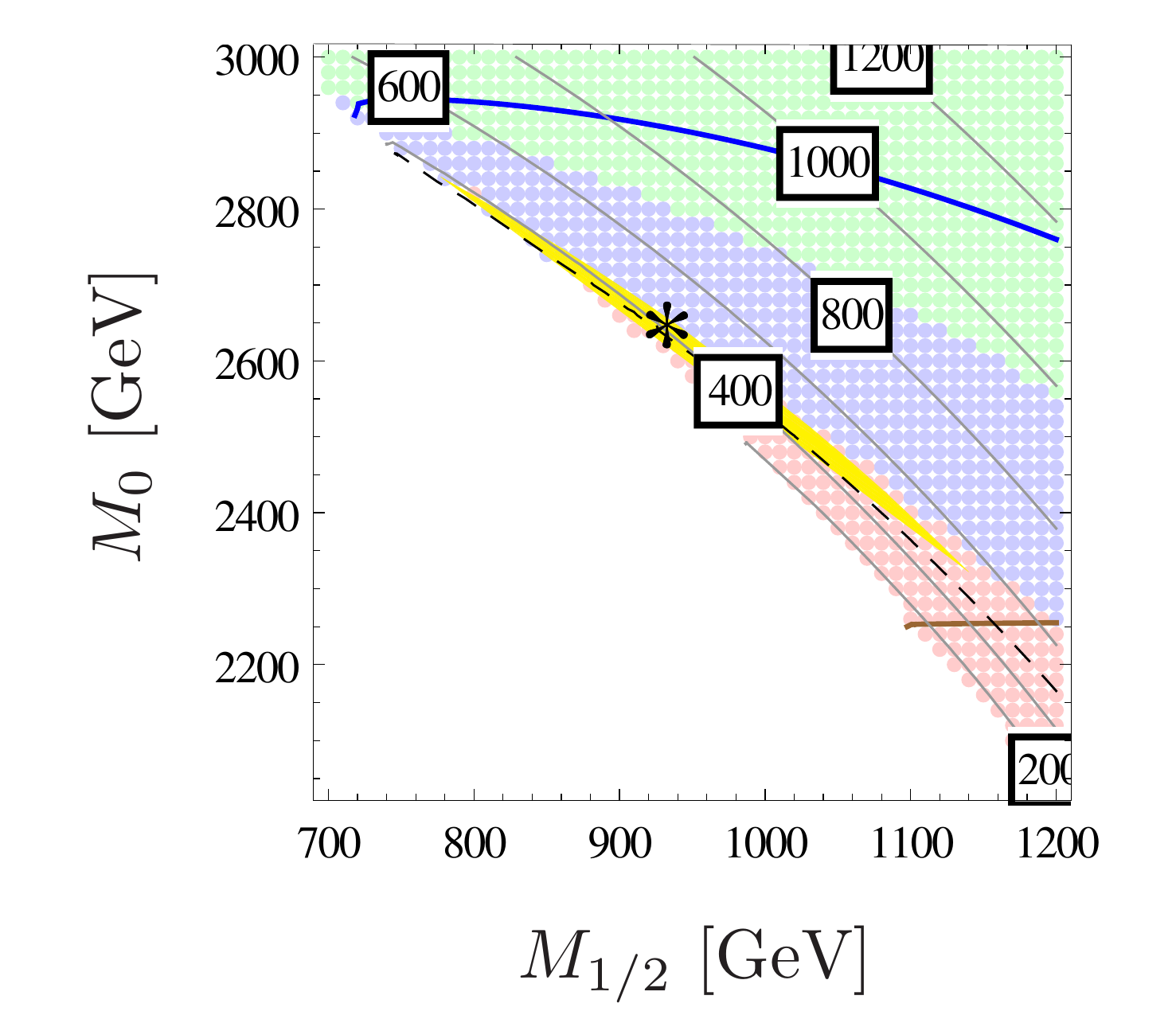} \hfill
\includegraphics[width=0.46\linewidth]{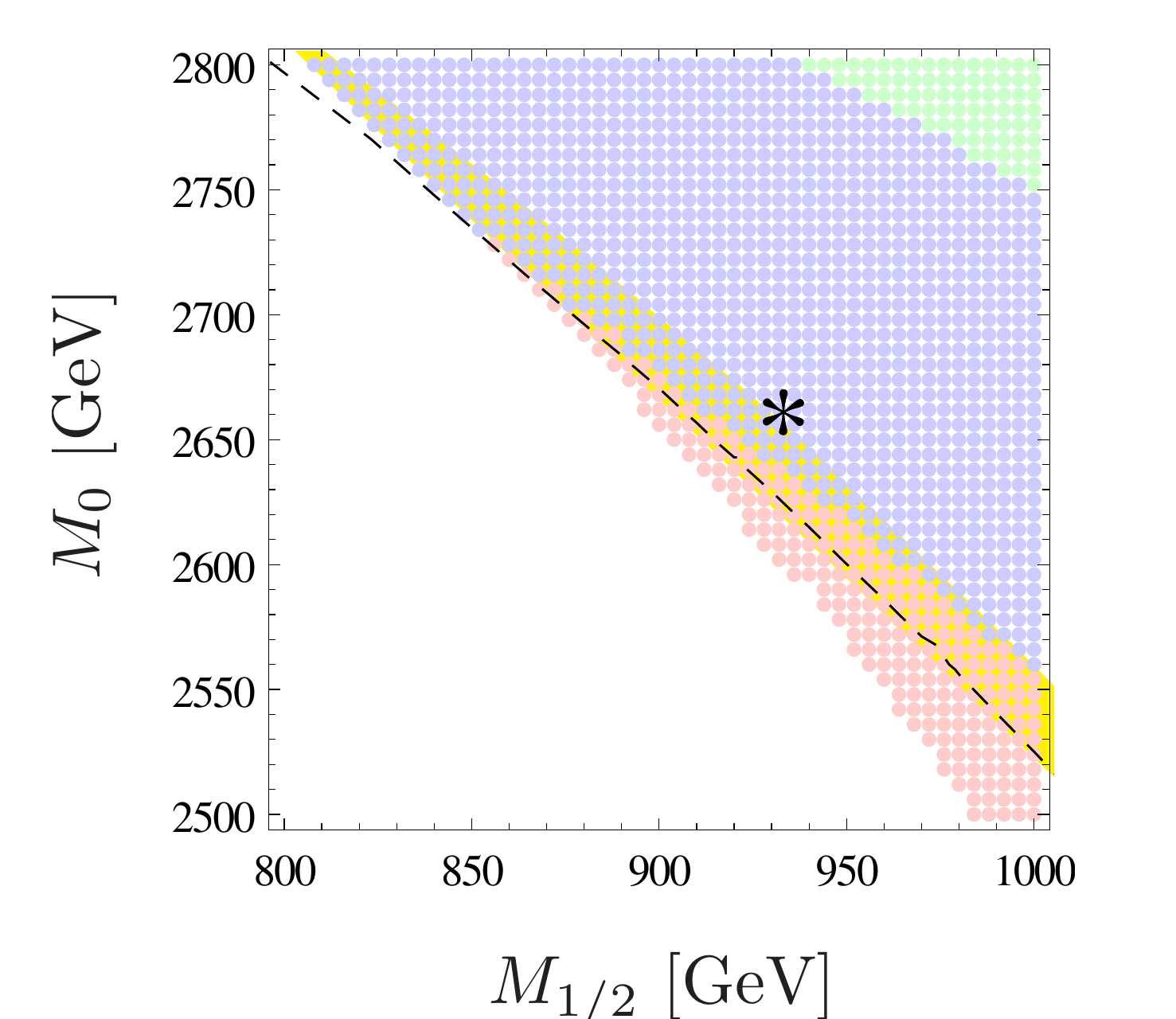}  
\caption{Mass of the light stop and vacuum stability in the $(M_{1/2}, M_0)$
 plane with $A_0=-6444$~GeV, $\mu<0$ and $\tan\beta=8.52$. The dashed line shows
 the transition to a charged LSP (neutralino LSP to the right of the line). The
 color coding is as in \Fig~\ref{fig:stopstau} and in \Sec~\ref{sec:rules}:
 points below the solid lines fail the corresponding conditions. In the yellow
 region, $\Omega h^2$ is in agreement with dark matter constraints (as in
 \Fig~\ref{fig:fittino_DM}). The star indicates the benchmark point 5.1 of
 \Ref~\cite{Cohen:2013kna}. The reason that this point is not in the strip with
 the correct dark matter abundance is that different SM input parameters were
 used in \Ref~\cite{Cohen:2013kna} in comparison to \Eq{\ref{eq:SMinputs}}:
 $m_t = 174.3$~GeV, $\alpha_S = 0.1172$ and $m_b(m_b)=4.25$~GeV. The line
 showing the division between passing and failing
 condition~\ref{eq:stop_large_tb_condition} using $0.65^{2}$ does not appear on
 this plot, but over-zealously would exclude the entire region shown, while
 taking $0.65^{2}$ would only exclude every point with
 $M_{0} \lesssim 2900$ \gev.}
\label{fig:dragon_DM}
\end{figure}

In addition to offering a new particle accessible to the LHC possibly motivated
 by naturalness arguments \cite{Barbieri:1987fn,Kitano:2006gv}, a stop NLSP
 could also give the correct neutralino abundance to explain the dark matter
 relic density. Even though stop co-annihilation usually works very efficiently
 for a larger mass splitting between the NSLP and LSP compared to the case for
 staus, the stops still have to be sufficiently light \cite{Ellis:2001nx}.
 Recent benchmark points for this scenario have been proposed in
 \Ref~\cite{Cohen:2013kna}. A scan arround their benchmark point 5.1
 ($M_0 = 2667$~GeV, $M_{1/2} = 933$~GeV, $\tan\beta=8.52$, $\mu<0$,
 $A_0 = -6444$~GeV) in the $(M_{1/2}, M_0)$ plane is shown in
 \Fig~\ref{fig:dragon_DM}, where we give the contour lines of the stop mass as
 well as the vacuum stability. The entire region which contains the stop
 coannihilation region has deeper CCB vacua and at least half of the points
 would tunnel from the DSB vacuum to CCB minima with an unacceptably short
 tunneling time. Note this can be a severe problem not only for stop
 coannihilation in the CMSSM, but also for pMSSM benchmark scenarios such as
 those discussed in \Ref~\cite{Harz:2012fz}. One solution to resurrect stop
 co-annihilation would be to go to much larger mass spectra; \eg the BP 5.2 of
 \Ref~\cite{Cohen:2013kna} with $m_{{\tilde{\chi}}^{0}_{1}} = 1$~TeV seems to be
 stable against tunneling to CCB minima.

\begin{figure}[tbp]
\centering
\includegraphics[width=0.45\linewidth
]{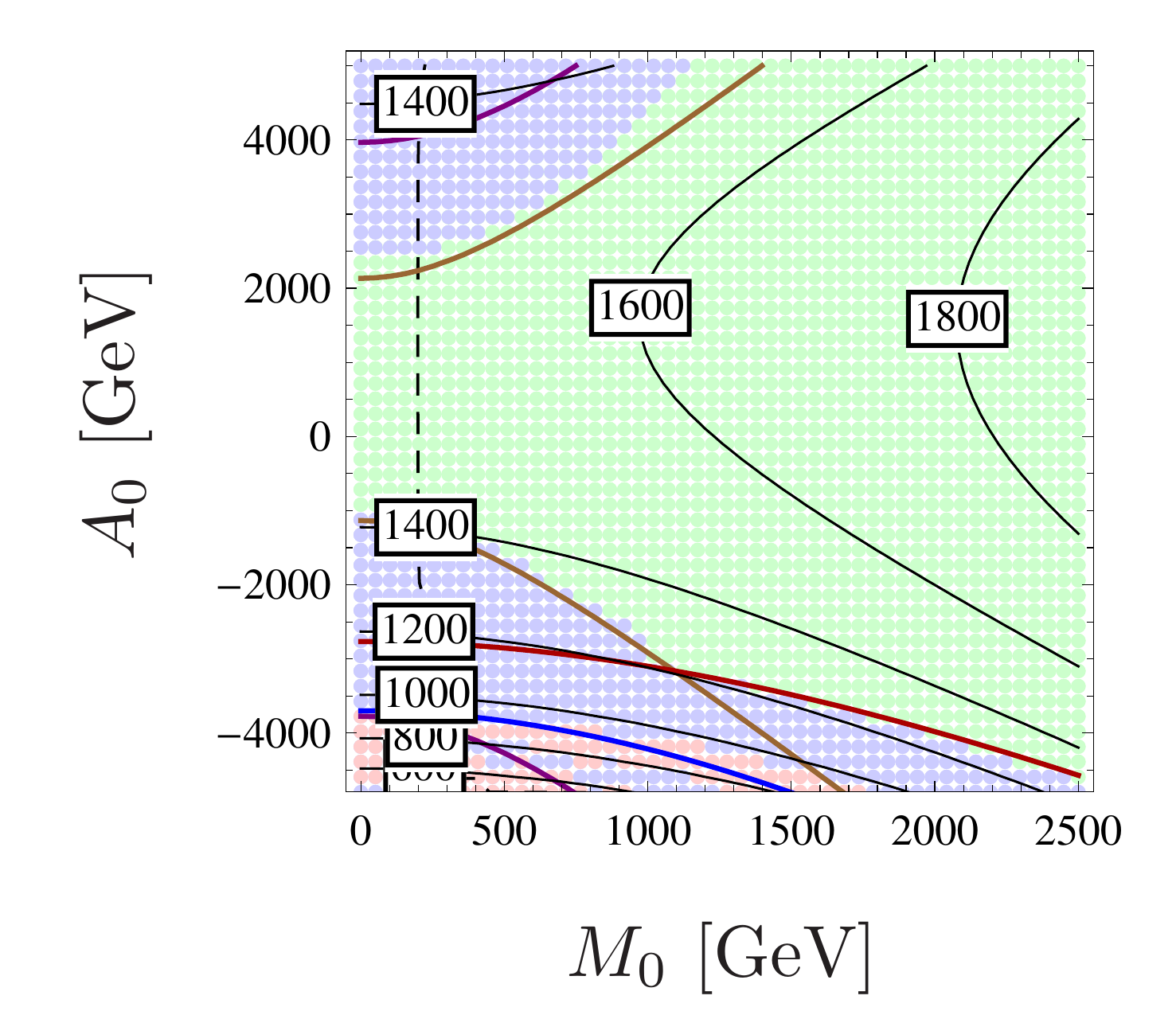}  \hfill 
\includegraphics[width=0.45\linewidth
]{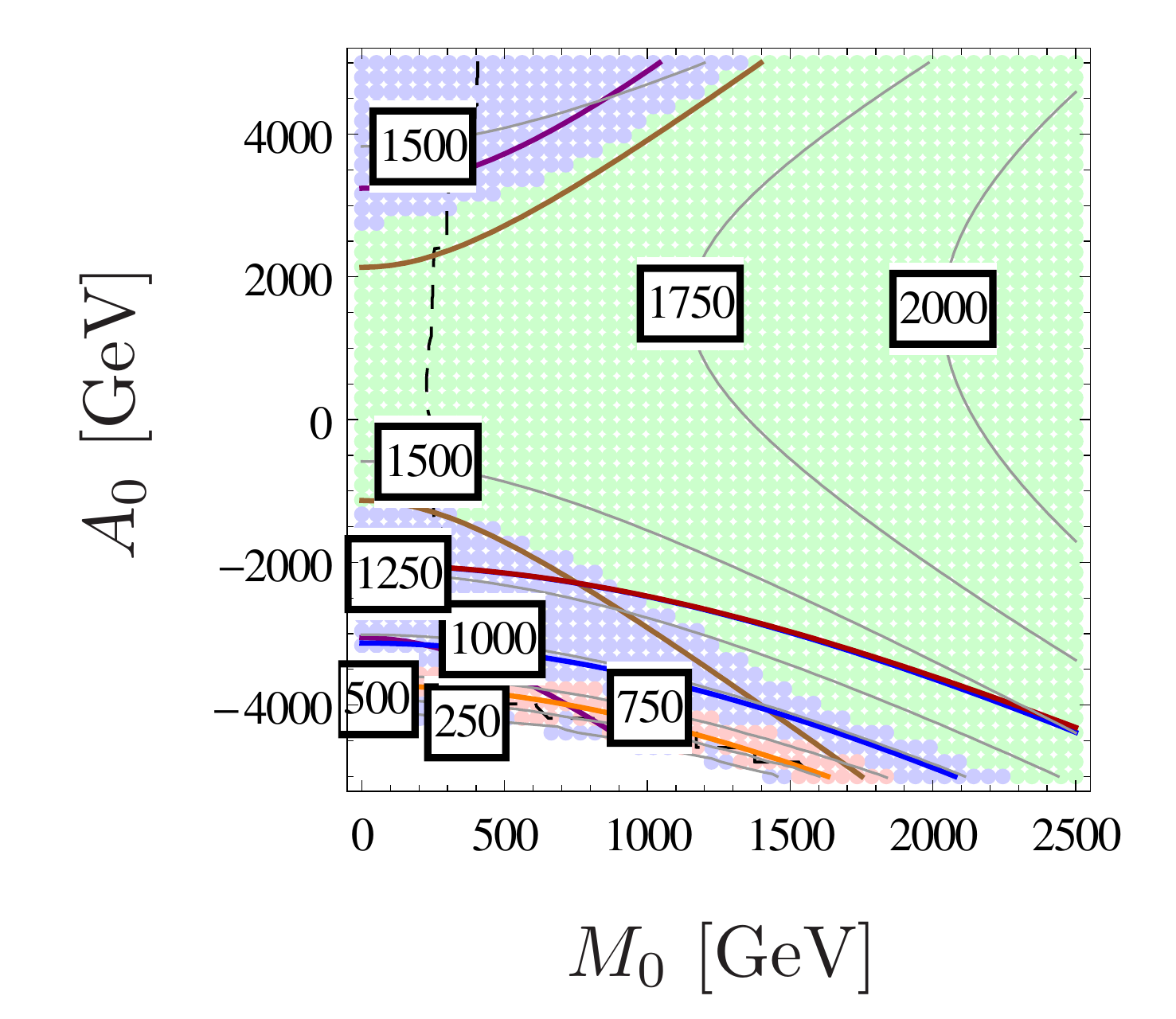} 
\caption{Stop mass and vacuum stability in the $(M_0, A_0)$ plane. We used
 $M_{1/2} = 1$~TeV, $\mu>0$ and $\tan\beta=2$ (left) or $\tan\beta=10$ (right).
 The color coding is as in \Fig~\ref{fig:stopstau} and in \Sec~\ref{sec:rules}:
 points to the left of the solid lines fail the corresponding conditions.
 Again, there is a degeneracy between the more exclusive blue line of
 condition~(\ref{eq:stop_large_tb_condition}) with the dark red line of
 condition~(\ref{eq:useless_sup_condition}). Points to the left of the dashed
 line also have charged LSPs.}
\label{fig:stop_vs_stability}
\end{figure}

We conclude this section with a comparison between our numerical results and
 the thumb rules given in the literature. We show in
 \Fig~\ref{fig:stop_vs_stability} the mass of the light stop and the vacuum
 stability in the $(M_0, A_0)$ plane for $M_{1/2}=1$~TeV and $\tan\beta=2$ or
 10. As can be easily seen, the conditions are rather irrelevant except for
 misusing conditions~(\ref{eq:useless_sup_condition})
 and~(\ref{eq:useless_GUT_condition}).

In general, the condition of stable or at least long-lived vacua in the case of
 small $\tan\beta$ and for negative $A_0$ puts severe limits on the mass of the
 light stop, and rapid tunneling times can rule out light stops in the CMSSM
 depending on the value of $M_{1/2}$.

\subsection{Constraining the parameter space of $m_h \simeq 125$~GeV}
\label{sec:Higgs}

The measurement of $m_h \simeq 125.5$~GeV~\cite{Aad:2012tfa, Chatrchyan:2012ufa}
 implies a powerful constraint on the parameter space of supersymmetric models
 due to the need for large radiative corrections. To get loop contributions
 which are large enough to increase the tree-level Higgs mass of
 $m^{(TL)}_h \leq m_Z$ by more than 30~GeV, one needs a large mass splitting in
 the stop sector, \ie $|A_0|$ must usually be large in comparison to $M_0$, see
 \Refs~\cite{Haber:1990aw, Haber:1996fp, Carena:2002es, Djouadi:2005gj,%
 Frank:2006yh} and references therein. At the one-loop level, the corrections
 are maximized for $X_t=A_t - \mu/\tan\beta \simeq  \sqrt{6} M_S$ where $M_S$ is
 the average stop mass. Taking two-loop corrections into account, this gets
 shifted to $X_t \simeq 2 M_S$ \cite{Carena:2000dp}.

However, as we have seen in the last sections, these are exactly the regions in
 the CMSSM parameter space which often suffer from a CCB minimum deeper than the
 DSB minimum. To demonstrate this, we take parameter planes discussed previously
 and show the calculated mass of the lightest Higgs boson. While the vacuum
 stability was calculated using the one-loop effective potential, these Higgs
 masses are based on a full diagrammatic one-loop calculation including the
 effects of the external momenta \cite{Pierce:1996zz} and, in addition, the
 known two-loop corrections are included
 \cite{Degrassi:2001yf, Brignole:2002bz, Brignole:2001jy, Dedes:2002dy}. For
 completeness, we also show the mass of the lighter Higgs boson evaluated at
 one-loop order beside each plot.

One might wonder about whether two-loop effects are important for the
 evaluation of the stability of the DSB vacuum, given that they are critical for
 obtaining the experimental value of the mass of the Higgs boson in the CMSSM.
 Unfortunately \vcs is currently restricted to one-loop order, and only the
 leading two-loop corrections at the DSB vacuum in the special case of the MSSM
 are known, let alone the full two-loop expression for arbitrary field
 configurations. However, given that the two-loop corrections to the Higgs
 mass-squared are of the order of $10\%$ compared to the one-loop corrections
 being of order $100\%$, along with the indications from
 \Fig~\ref{fig:Q_dependence} that loop corrections play a subdominant role if
 the scale is chosen judiciously, as expected, as the usual SUSY scale, it would
 seem that the loop expansion converges well and the higher orders should not
 affect the results substantially.

\begin{figure}[tbp]
\centering
\includegraphics[width=0.49\linewidth
]{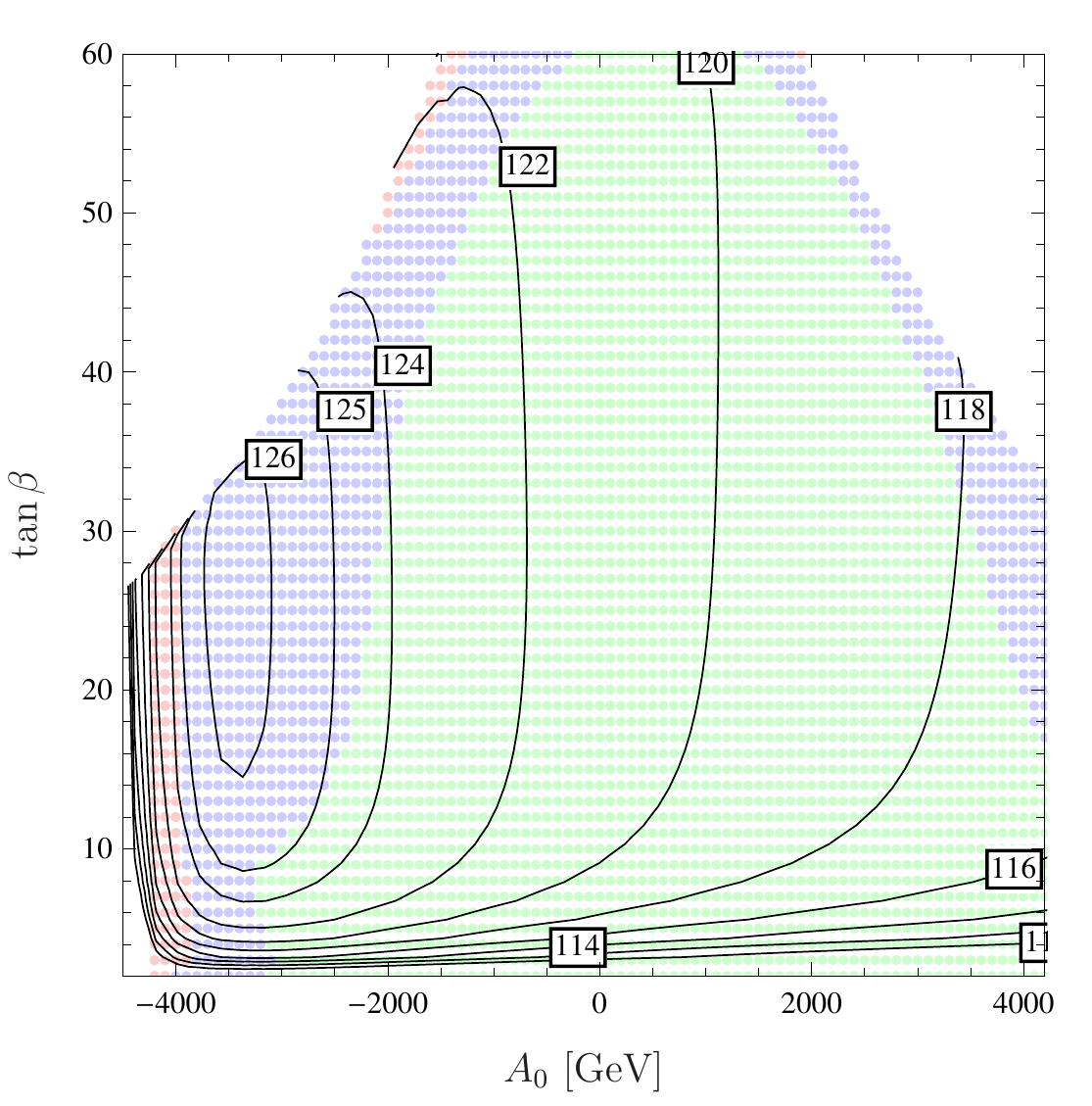}  \hfill 
\includegraphics[width=0.49\linewidth
]{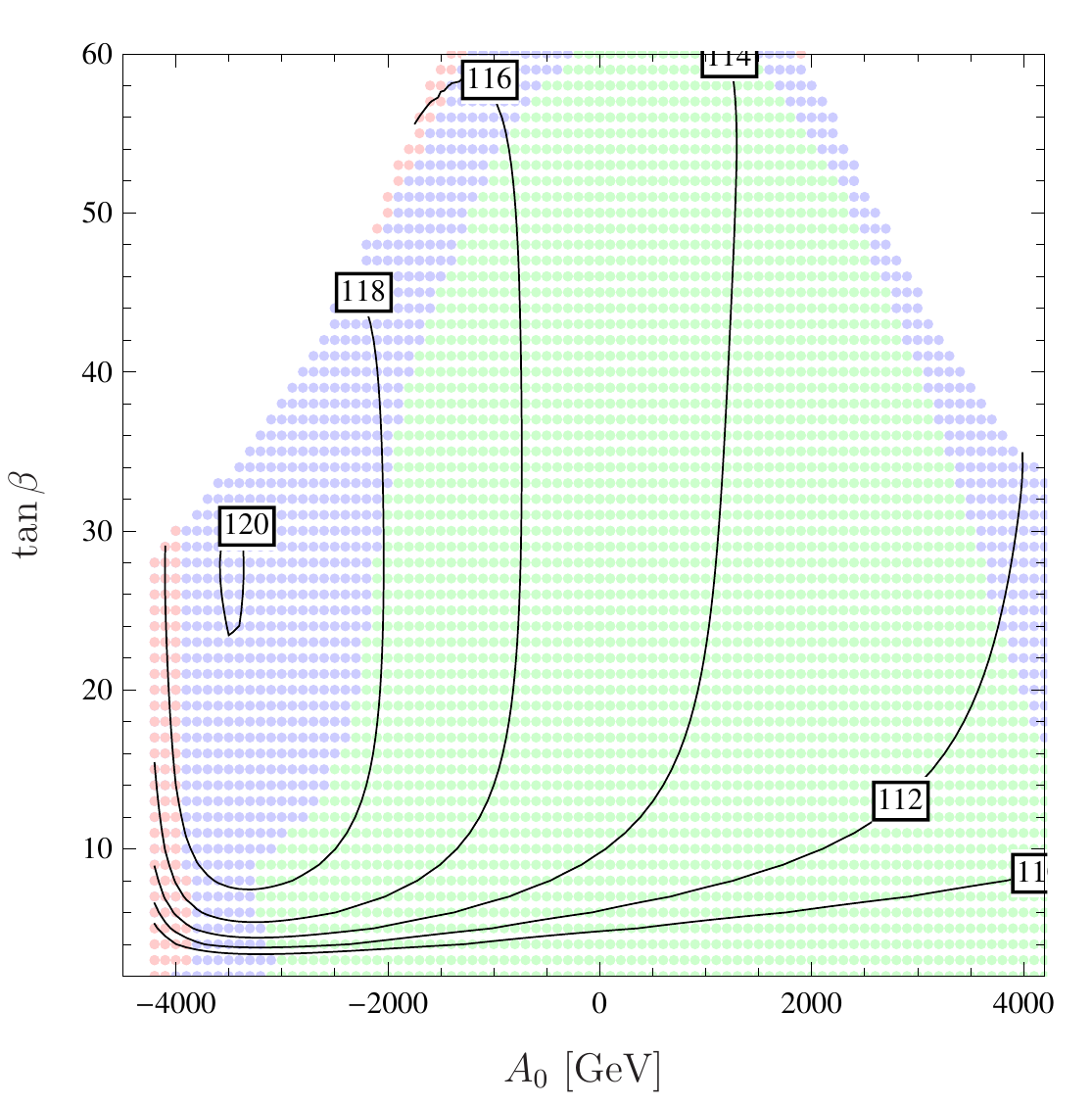}
\caption{Vacuum stability and the Higgs mass in the $(A_0, \tan\beta)$ plane for
 fixed $M_0 = M_{1/2} = 1$~TeV and $\mu>0$. On the left, the leading two-loop
 corrections at the DSB minimum have been taken into account, while on the
 right, the mass of the lighter Higgs boson is shown only at one-loop order. The
 color code is the same as in \Fig~\ref{fig:stopstau}.}
\label{fig:HiggsMass_vs_Stability_A0_tb}
\end{figure}

To show how serious the Higgs mass constraint in combination with vacuum
 stability is, we start with moderate SUSY masses and put
 $M_0 = M_{1/2} = 1$~TeV. The Higgs mass and the vacuum stability in the
 $(A_0, \tan\beta)$ plane is shown in
 \Fig~\ref{fig:HiggsMass_vs_Stability_A0_tb}. We see that nearly the entire line
 where $m_h = 125$~GeV is located in a range with charge- or color-breaking
 minima deeper than the DSB minima.

\begin{figure}[tbp]
\centering
\includegraphics[width=0.45\linewidth
]{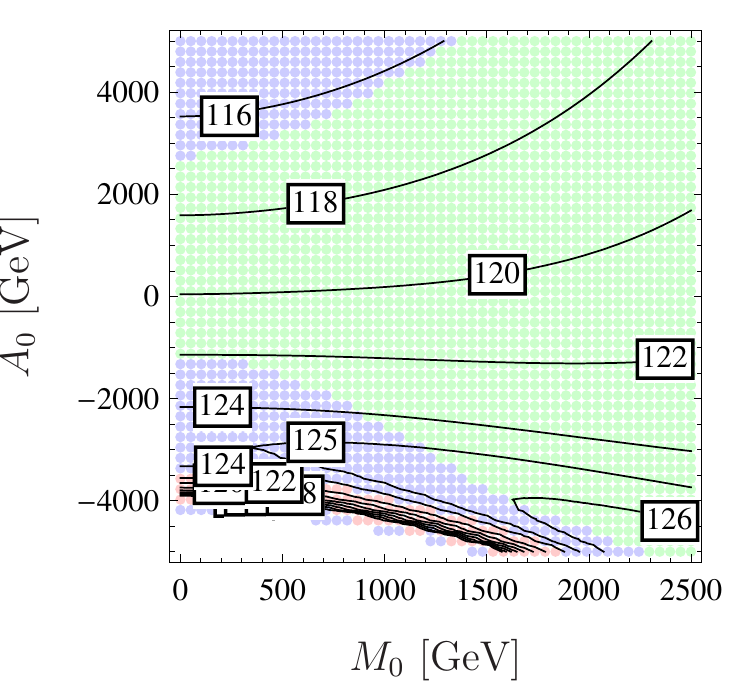}
\hfill
\includegraphics[width=0.45\linewidth
]{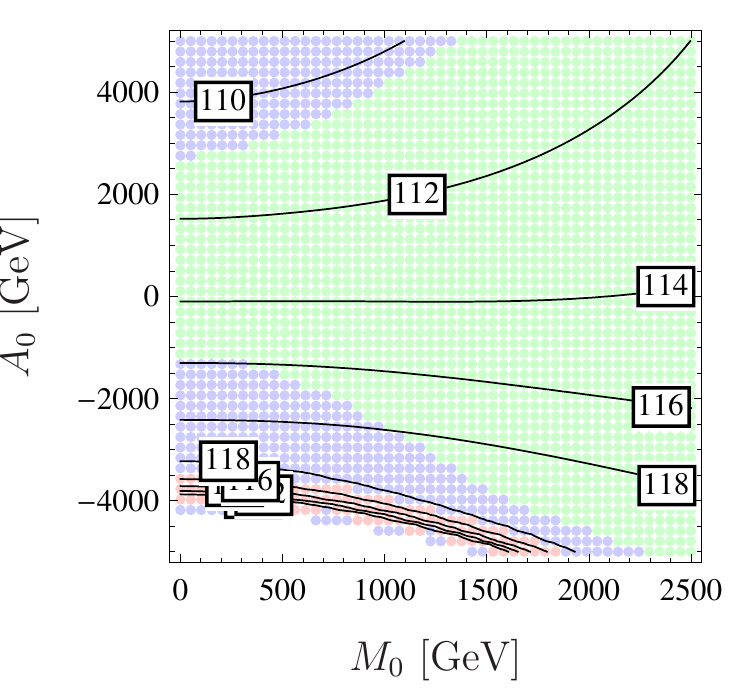} \vspace{0.3cm}\\
\includegraphics[width=0.45\linewidth
]{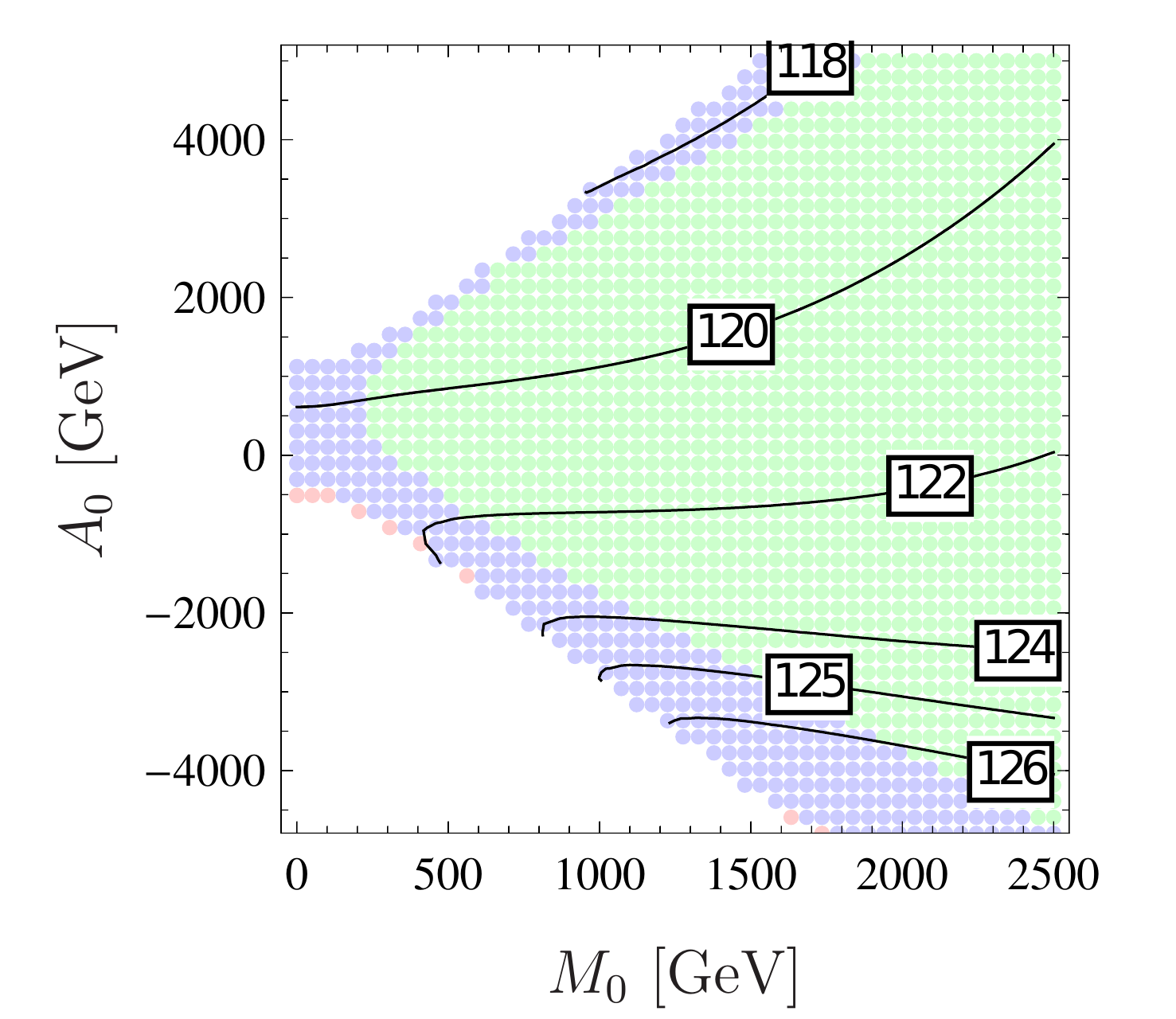}
\hfill
\includegraphics[width=0.45\linewidth
]{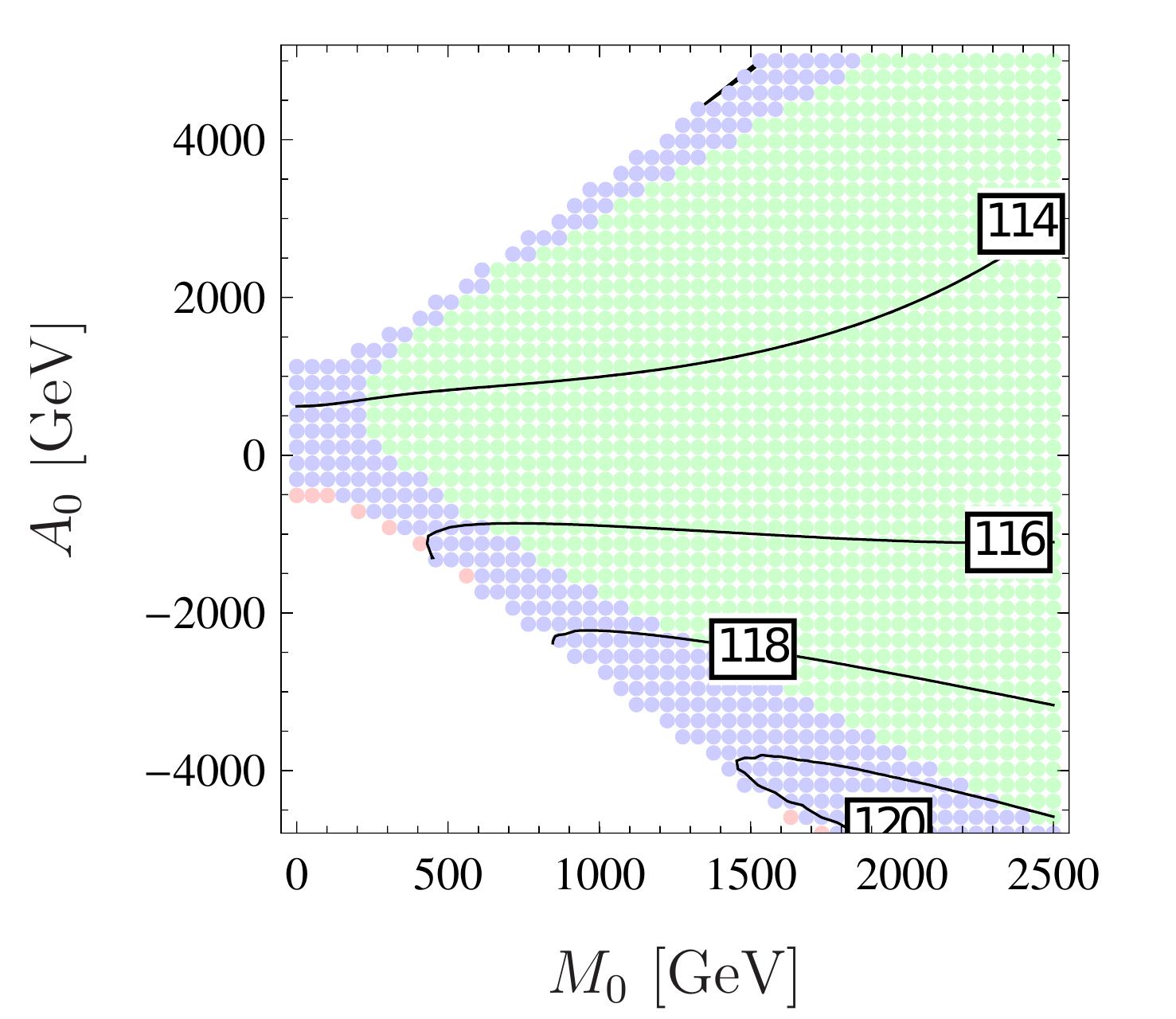} 
\caption{Vacuum stability and the Higgs mass in the $(M_0, A_0)$ plane for fixed
 $M_{1/2} = 1$~TeV, $\mu>0$ and $\tan\beta=10$ (upper) or $\tan\beta=40$
 (lower). On the left, the leading two-loop corrections to the mass of the
 lighter Higgs boson have been taken into account, on the right, only the mass
 at one-loop order is shown. The color code is the same as in
 \Fig~\ref{fig:stopstau}.}
 \label{fig:HiggsMass_vs_Stability_m0_A0}
\end{figure}

To test if the tension between the vacuum stability and the Higgs mass relaxes
 in the case of heavier SUSY spectra, we consider in
 \Fig~\ref{fig:HiggsMass_vs_Stability_m0_A0} the $(M_0, A_0)$ plane and allow
 $M_0$ up to $2.5$~TeV. In addition, we used $M_{1/2} = 1$~TeV, $\mu>0$ and
 $\tan\beta=10, 40$. For both values of $\tan\beta$, the correct Higgs mass can
 only be reached for large negative values of $A_0$, as expected.
 Especially for small $M_0$, the ratio $|A_0/M_0|$ must be huge to push the
 Higgs mass to the desired value. As a consequence, all points with
 $m_h>124$~GeV are in an area with CCB minima deeper than the DSB minima for
 $M_0<900$~GeV ($\tan\beta=10$) respectively $M_0<1200$~GeV ($\tan\beta=40$).
 The lower bound on $M_0$  increases by about 300~GeV if one demands
 $m_h \geqslant 125$~GeV. Only for larger $M_0$ it is possible to find points
 with the Higgs mass in the correct range and a DSB vacuum stable against
 tunneling to CCB vacua.

\section{Discussion and conclusion}

As we have shown, the calculation of tunneling times from a phenomenological
 minimum to charge- and/or color-breaking minima can be taken as a new
 constraint for excluding unphysical parameter points.

A lot of effort has been spent in the literature to find best-fit points that
 consider a vast amount of phenomenology data. However, vacuum stability is
 either not considered or introduced by the use of one or more of the conditions
 which we have shown to be obsolete. This is not limited to the CMSSM: for
 instance, the general MSSM points given by the `natural SUSY' benchmark of
 \cite{Baer:2012yj} and benchmark `I' of \cite{Harz:2012fz} suffer from
 unacceptably short tunneling times to CCB minima with stop \vevs.

We have discussed the impact of the vacuum stability on the parameter space of
 the CMSSM. The difference to previous studies is that we have used a fully
 numerical approach ensuring that all deeper charge- and color-breaking minima
 at tree level are found. In addition, the inclusion of the one-loop corrections
 to the effective potential reduced the uncertainty from the scale at which the
 potential is evaluated. It turned out that approximate thumb rules which have
 been analytically derived in the literature do not exclude much of the relevant
 parameter space where the DSB vacuum suffers from a rapid decay into a deeper
 minimum with broken $U(1)_{EM}$ and maybe even $SU(3)_{c}$. This happens
 especially in areas of the parameter space with a large stau and/or stop mass
 splitting. However, these are the regions which have been considered as
 phenomenlogically preferred since they could explain the abundance of dark
 matter due to a co-annihilation mechanism. It has been shown that many regions
 discussed in this context are already ruled out if one includes the constraint
 that the DSB vacuum must be stable or at least long-lived metastable.

In addition, the large stop mass splitting is also needed to explain the
 measurement of the Higgs mass within the CMSSM. We found that for SUSY masses
 in the lower TeV range hardly any region in the CMSSM can be found which leads
 to a Higgs mass of 125~GeV or above without suffering from CCB global minima.

\section*{Acknowledgments}
The authors would like to thank Sasha Belyaev for useful discussions on the
 limits of the stau co-annihilation strip. This work has been supported by the
 DFG, research training group GRK 1147 and project No.\ PO-1337/2-1. FS is
 supported by the BMBF PT DESY Verbundprojekt 05H2013-THEORIE ``Vergleich von
 LHC-Daten mit supersymmetrischen Modellen''.

\bibliography{CMSSM_CCB.bib}
\bibliographystyle{JHEP}

\end{document}